\documentclass[usenatbib]{mnras}

\usepackage{newtxtext,newtxmath}

\usepackage[T1]{fontenc}

\usepackage[utf8]{inputenc}
\usepackage[T1]{fontenc}
\usepackage{geometry}
\usepackage{amsmath, amsfonts} 
\usepackage{graphicx}
\usepackage{titlesec}

\usepackage{setspace}
\usepackage{enumitem}
\usepackage{natbib}
\usepackage{hyperref}
\usepackage{orcidlink}

\newcommand{\thesis}[1]{} 

\hypersetup{
    colorlinks=true,
    linkcolor=blue,
    filecolor=blue,      
    urlcolor=blue,
    citecolor=blue ,
    pdftitle={Overleaf Example},
    pdfpagemode=FullScreen,
}

\newcommand{\ur}{u_r}
\newcommand{\csg}{\langle c_{\rm s}\rangle_g}

\newcommand{\kms}{\,{\rm km}\,{\rm s}^{-1}}

\newcommand{\sigmat}{\sigma_{\rm t}}
\newcommand{\Mhalo}{M_{\rm halo}}

\newcommand{\tdiss}{t_{\rm diss}}
\newcommand{\tflow}{t_{\rm flow}}
\newcommand{\Rcirc}{R_{\rm circ}}
\newcommand{\Rsonic}{R_{\rm sonic}}
\newcommand{\nH}{n_{\rm H}}
\newcommand{\msun}{\,{\rm M}_\odot}
\newcommand{\tcools}{ {t}_{\mathrm{cool}}^{({\mathrm{s}})}}
\newcommand{\tcool}{ {t}_{\mathrm{cool}}}
\newcommand{\tff}{ {t}_{\mathrm{ff}}}
\newcommand{\vc}{V_{\rm c}}
\newcommand{\cs}{c_{\rm s}}
\newcommand{\Rvir}{R_{\rm vir}}
\newcommand{\Tvir}{T_{\rm vir}}

\newcommand{\uin}{|\ur|}

\def\apjl{ApJ}%
\def\apjs{ApJS}%
\def\ao{Appl.~Opt.}%
\def\aap{A\&A}%
\title[Accretion-Driven CGM Turbulence]{Accretion-Driven Turbulence in the Circumgalactic Medium}

\author[R.~Goldner et al.]{
\parbox{\textwidth}{
 Roy Goldner\,\orcidlink{0009-0006-0305-7151},$^{\dagger}$\thanks{E-mail: roy.goldner@gmail.com; sternjon@tauex.tau.ac.il}
 Jonathan Stern\,\orcidlink{0000-0002-7541-9565},$^{1}$
 Drummond Fielding\,\orcidlink{0000-0003-3806-8548}$^{2}$,
Claude-Andr\'e Faucher-Gigu\`ere\,\orcidlink{0000-0002-4900-6628}$^{3}$,
Yakov Faerman\,\orcidlink{0000-0003-3520-6503}$^{1}$ and 
Aharon Kakoly\,\orcidlink{0009-0001-9850-8827}$^{1}$
}
\vspace{0.4cm}\\
\parbox{\textwidth}{
 $^{1}$School of Physics and Astronomy, Tel Aviv University, Tel Aviv 69978, Israel \\
 $^2${Department of Physics, New York University, 726 Broadway, New York, NY 10003, USA}\\
$^3${Department of Physics \& Astronomy and CIERA, Northwestern University, 1800 Sherman Ave, Evanston, IL 60201, USA}\\
}}
\vspace{-40pt}

 \date{Accepted XXX. Received YYY; in original form ZZZ}

\pubyear{\the\year}

\newcommand{\changed}[1]{#1}

\begin{document}
\label{firstpage}
\pagerange{\pageref{firstpage}--\pageref{lastpage}}

\maketitle

\begin{abstract}
Simulations suggest that turbulence is ubiquitous in the circumgalactic medium (CGM), \changed{though the source and properties of CGM turbulence is uncertain. Using analytic considerations and hydrodynamic simulations we study how CGM turbulence is driven by gas accretion,  
  thus providing a baseline for additional turbulence driving processes such as galaxy feedback. 
We demonstrate that in halos with mass  $\sim10^{10}-10^{12}\msun$ at $0< z < 2$, accretion amplifies mild turbulent velocities near the virial radius of $\sigmat(\Rvir)\sim10\kms$ to virial velocities at inner CGM radii, $\sigmat(0.1\Rvir)\approx v_{\rm vir}\sim100\kms$.} Rapid cooling at these inner radii further implies that thermal pressure support is small, and the gas is dominated by the cool and warm ($\sim10^4-10^5\,{\rm K}$) phases.   
Inner CGM energetics in these halos is thus dominated by turbulence, with gas density distributions and velocity structure functions  similar to those seen in simulations of \textit{isothermal} supersonic turbulence, rather than those seen in subsonically turbulent stratified media such as the ICM. 
The accretion rate in these systems is regulated by the turbulence dissipation rate, \changed{in contrast with being regulated} by the cooling rate as in more massive halos. 
We argue that galaxy feedback is unlikely to qualitatively change our conclusions unless it \changed{significantly depletes the CGM} or continuously injects high specific energy material ($\gg v^2_{\rm vir}$). 
Such `turbulence-dominated' CGM can be identified in observations via the predicted wide lognormal ionization distributions and large velocity dispersions in UV absorption spectra. 
\end{abstract}



\section{Introduction}
The interaction of galaxies with their environment via accretion and outflows is mediated by the circumgalactic medium (CGM), loosely defined as the gas extending from the edge of the galactic disk to the halo virial radius ($\sim10-300\,{\rm kpc}$ for the Milky-Way CGM). Understanding CGM dynamics and properties is therefore essential for understanding galaxy formation and evolution \citep{Tumlinson17, FaucherGiguere23}. Due to its high Reynolds number the CGM is expected to be turbulent, a prediction \changed{consistent with} several observations. These include non-thermal broadening of circumgalactic absorption profiles both in the UV \citep[e.g.,][]{Rauch96, Rauch01, Werk16, Rudie19, Chen23} and potentially in the X-ray \citep{Gupta12,Fang15,Miller16,Faerman17}, the velocity difference between multiple absorption features along a given line of sight and relative to the central galaxy \citep[e.g.,][]{Werk13,Borthakur16, Huang16}, and indications of low thermal pressure in the cool $\sim 10^4\,{\rm K}$ CGM which could indicate turbulent pressure support \citep[see discussion in][]{Faerman23}. \changed{A recent analysis by \cite{Li26a,Li26b} of absorption and emission line spectra in local starburst galaxies also point to the importance of turbulence in the inner CGM.}

Previous studies of turbulence in halo gas have focused mainly on cluster scale halos in which the volume-filling phase is more observable, while CGM turbulence has been less systematically explored. The driving of turbulence in the intracluster medium (ICM) is typically attributed to jet-inflated bubbles and galaxy mergers \citep[e.g.][]{Churazov02, Omma04,Gaspari18, Mohapatra19}. 
Although it is tempting to treat the CGM as a scaled-down version of the ICM, several properties of the CGM suggest that there are qualitative differences, rather than just quantitative differences, between the two systems. First, the CGM does not have a `cooling flow problem', i.e.~the CGM X-ray luminosity in the Milky-Way and nearby massive disks is comparable to that expected based on their star formation rates \citep{Miller15,Bogdan17,Das19,stern19}, in contrast with a discrepancy of $1-2$ orders of magnitude in the ICM \citep{McDonald18}. 
Second, cooling times are generally shorter in the CGM than in the ICM, and specifically can be shorter than dynamical times below the Milky-Way mass scale \citep{white78, birnboim03, Fielding17, stern20}. \cite{Fielding17} demonstrated that such short cooling times cause accretion and outflow shocks to be highly radiative, so kinetic energy dominates over thermal energy and CGM turbulence is largely supersonic. A supersonically turbulent CGM would have gas density distributions and velocity structure functions which differ qualitatively from the subsonically turbulent ICM.  
Thus, identifying at which masses, redshift and radii turbulence is supersonic rather than subsonic is crucial for our understanding of the CGM. 
Supersonic CGM have also been identified in cosmological simulations, mainly in the inner CGM of $\sim L^\star$ galaxies at redshifts of $z\sim0.5$ or higher and in the CGM of dwarf galaxies, both in FIRE \citep{Stern21A,Stern21B,Li21,Gurvich23,Kakoly25,Sun25} and potentially in IllustrisTNG \citep{Semenov24}.

In this paper we study the properties of CGM turbulence using analytic considerations and idealized 3D hydrodynamic simulations, focusing on the roles of gas accretion and the dependence on typical gas cooling times. Previous idealized studies have either studied how turbulence arises from the interaction of accretion and feedback \citep{Fielding17,Lochhaas20}, or have assumed an unspecified external driving mechanism and studied its implications on CGM properties as a function of turbulence parameters \citep{Buie20,Buie22,Tan21,Gronke22, Koplitz23, Das24, Lv24}.  Our choice of focusing on accretion driving of turbulence, which has relatively constrained parameters, allows us to provide a robust baseline for CGM turbulence, onto which the effects of additional processes can be added. We also demonstrate that considering accretion only is sufficient to capture the qualitative transition between supersonic and subsonic CGM turbulence as the halo grows in mass. 

To study how accretion drives CGM turbulence we utilize two distinct analytic frameworks: the theory of how general accretion flows drives turbulence, presented by \cite{Robertson12} and further explored in the context of star forming clouds \citep{Murray15,Murray17,Hennebelle21}; and the theory of spherical CGM inflows from  \cite{stern19,stern20,Stern24} which is based on cooling flow theory developed for the ICM \citep{mathews78, fabian84, Bertschinger89,balbus89}. 
These two components are combined in section~\ref{sec:analytic} to derive new idealized solutions for turbulent CGM inflows. In sections~\ref{sec:setup} and \ref{sec:results} we present and analyze our 3D hydrodynamic simulations, with which we both test the analytic results and examine statistical properties of CGM turbulence such as velocity structure functions (VSFs) and gas density distributions. We discuss the implications and caveats of our results in section~\ref{sec:discussion} and summarize in section~\ref{sec:summary}.

\section{Analytic Framework}\label{sec:analytic}
In this section, we outline the analytic framework for our study. 
We begin by deriving how accretion drives turbulence according to the \cite{Robertson12} model in section~\ref{sec:turb} and reviewing the \cite{stern19,stern20} solutions for spherical CGM inflows based on cooling flow theory in section~\ref{sec:inflows}. We then combine these two frameworks to deduce the properties of accretion-driven CGM turbulence, and the back reaction of turbulence on CGM accretion (section~\ref{sec: turb inflows}). The implied density distributions and velocity structure functions in the CGM are derived in sections~\ref{sec:density_distribution} -- \ref{sec:VSF analytic}.

\subsection{Accretion-driven turbulence}\label{sec:turb}

In this section we discuss the transfer of energy from the mean flow to turbulent fluctuations in accreting gas. The derivation follows that of \cite{Robertson12}, \cite{Murray15}, \cite{Murray17}  and \citet{Hennebelle21}.

We characterize the velocity field of an accreting CGM as $\vec{u}(r,\Omega,t)$, where $r$ is the distance from the halo center, $\Omega$ is the solid angle and $t$ is time. For an accretion flow which is statistically spherical and time-steady, we can separate $\vec{u}$ into a mean radial component and a fluctuating component, where the mean radial speed and the fluctuation statistics depend solely on radius $r$. 
The mass-weighted mean velocity is thus
\begin{equation}
    \langle \vec{u}\rangle_{\Omega,t}
    \equiv\frac{\iint \rho\vec{u}d\Omega dt}{\iint \rho d\Omega dt} \equiv u_r(r)\hat{r} ~,
\end{equation}
while the fluctuating component of the velocity field is characterized by the 3D turbulent velocity $\sigmat(r)$, defined as
\begin{equation}
    \sigmat^2(r)\equiv \left<{ \left( \vec{u}-\ur\hat{r}\right)^2}\right>_{\Omega,t}
    \label{eq:sigma_t} ~.
\end{equation}
\citet{Robertson12} demonstrated that $\sigma_{\rm t}$ increases in a contracting turbulent flow in a manner similar to adiabatic heating of thermal motions, a result familiar from the study of peculiar velocities in cosmology. For homologous contraction ($u_r\propto r$) they deduced the following `adiabatic heating' term for the evolution of $\sigma_{\rm t}$ (see derivation in Appendix~\ref{app:adiabatic heating}): 
\begin{equation}
    \left.\frac{d \sigmat}{dt}\right|_{\rm heating}=\sigmat \frac{{\uin}}{r}
	\label{eq:heating}
\end{equation}
where $d/dt$ is the Lagrangian derivative. Note that for quasi-steady conditions in which $\partial\sigmat/\partial t=0$ this relation is equivalent to $\sigmat r=\text{constant}$, and implies that the turbulent velocity scales as $\propto r^{-1}$ in the absence of other effects such as turbulence dissipation. 
Eqn.~(\ref{eq:heating}) can also be viewed as a transfer of kinetic energy between the converging scale $r$ and smaller scales (or the other way around in an outflow). Such inward and upward energy cascades associated with compression and expansion have been identified in simulations of compressible turbulence \citep{Wang12,Wang18,Tonicello22}. 

Quasi-spherical accretion does not necessarily satisfy $u_r\propto r$ as assumed by \cite{Robertson12}. \cite{Hennebelle21} relaxed this assumption by deriving the adiabatic heating term directly from the fluid equations, and showed that the factor $u_r/r$ in the right hand side of eqn.~(\ref{eq:heating}) is formally valid only for tangential velocity fluctuations, while it should be replaced with $du_r/dr$ for radial fluctuations. For a general non-homologous contraction the adiabatic heating term is thus anisotropic. However, in practice the tangential and radial velocity fluctuations are coupled, so in the analytic derivation we disregard this complication and keep using the isotropic adiabatic heating term from eqn.~(\ref{eq:heating}). In the numerical simulations below we assess the level of anisotropy due to this effect and due to other effects such as buoyancy \citep[e.g.,][]{Mohapatra20,Mohapatra21}.

In addition to the adiabatic heating term in eqn.~(\ref{eq:heating}), turbulence energy is expected to dissipate on
 a timescale $\tdiss$ which is similar to the turnover time of the largest eddies $t_{\mathrm{eddy},\ell_{\rm t}}$ \citep[e.g.,][]{maclow99}:
\begin{equation}
    \frac{1}{2}\left.\frac{d \sigma_\text{t}^2}{dt}\right|_{\rm dissipation}=-\eta \frac{{\sigma_\text{t}^2}}{ t_{\mathrm{eddy},\ell_{\rm t}}}=-\eta \frac{{\sigma_\text{t}}^3}{\ell_\text{t}}
	\label{eq:dissipation}
\end{equation}
where $\ell_\text{t}$ is the driving scale of the turbulence, $\eta$ is a constant of order unity known as the `dissipation constant', and we assumed $\sigmat^2$ is dominated by the largest eddies. 
The value of $\ell_\text{t}(r)$ in a contracting medium is no larger than the size of the shell $r$, and we show below that in most cases a CGM accretion flow will converge to $\ell_\text{t}\approx r$. We thus get for the dissipation term
\begin{equation}
    \left.\frac{d \sigma_\text{t}}{dt}\right|_{\rm dissipation}=-\eta\frac{{\sigma_\text{t}}^2}{r}  
	\label{eq:dissipation short}
\end{equation}
as also deduced by \cite{Murray15}. This relation can be equivalently stated by defining the dissipation timescale as
\begin{equation}\label{eq:tdiss}
\tdiss=\frac{r}{\eta\sigmat} ~.
\end{equation}

\begin{figure}
	\includegraphics[width=\columnwidth]{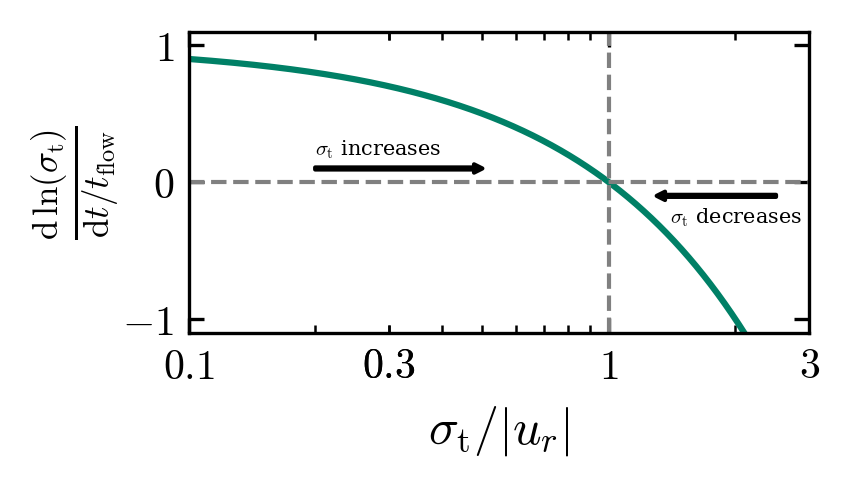}
\vspace{-0.6cm}
\caption{Lagrangian evolution of the turbulent velocity in inflows, versus the ratio of the turbulent velocity to the mean inflow velocity. Calculation is based on eqn.~(\ref{eq:turbulence_combine}) assuming $\eta=1$, taken from Robertson \& Goldreich (2012). The turbulent velocity increases when it is lower than $\uin$ and decreases when it is higher than $\uin$, converging to $\sigmat=\uin$.}
    \label{fig:partial_time_derivative}
\end{figure}

Combining the adiabatic heating term (eqn.~\ref{eq:heating}) and the dissipation term (eqn.~\ref{eq:dissipation short}) gives
\begin{equation}
    \frac{d \sigma_\text{t}}{dt}=\frac{\sigma_t}{r}\left(-u_r-\eta \sigma_\text{t}\right) ~,
	\label{eq:turbulence_combine base}
\end{equation}
and by defining also
\begin{equation}\label{eq:tflow}
    \tflow\equiv \frac{r}{|u_r|}
\end{equation}
we get 
\begin{equation}
    \frac{d \ln \sigmat}{dt/\tflow}= 1-\frac{\eta \sigmat}{|u_r|} ~.
	\label{eq:turbulence_combine}
\end{equation}
This relation is plotted in Figure~\ref{fig:partial_time_derivative} for $\eta=1$. The figure demonstrates that in an inflow $\sigmat$ converges, in a Lagrangian sense, to a value of $\eta^{-1}|u_r|$ at which the adiabatic heating is balanced by dissipation. If $\eta\sigma_t> |u_r|$ then dissipation dominates and $\sigmat$ decreases, while if $\eta\sigma_t< |u_r|$ then adiabatic heating dominates and $\sigma_t$ increases. 
Note also that for gas expanding in an outflow ($u_r>0$) both the adiabatic and dissipation terms in eqn.~(\ref{eq:turbulence_combine base}) are negative so turbulence necessarily decreases with the flow.

Equation~(\ref{eq:turbulence_combine base}) also demonstrates that a turbulent inflow is statistically steady if $u_r$ and $\sigmat$ evolve on timescales that are long compared to $t_{\rm eddy}$ and $\tflow$. Such a quasi-steady state is achievable especially in the inner CGM where dynamical and cooling timescales are much shorter than the Hubble time, on which CGM outer boundary conditions and the halo potential evolve. Under this quasi-steady assumption we can use $d/d t=u_r\partial/\partial r$ and thus eqn.~(\ref{eq:turbulence_combine base}) yields the following Eulerian relation between $\sigmat$ and $u_r$
\begin{equation}
    \frac{\sigmat}{\uin} =\frac{1+p}{\eta} ~,
	\label{eq:sigmat_ur}
\end{equation}
or equivalently
\begin{equation}\label{eq:tflow vs tdiss}
\tflow = (1+p)\tdiss ~,
\end{equation}
where we defined 
\begin{equation}
p\equiv \frac{\partial \ln\sigmat}{\partial\ln r}~.
\end{equation}
For a power-law solution where $p$ is constant and assuming $p>-1$, we get that $\sigma_t$ is proportional to $u_r$, so it also holds that $p=\partial \ln u_r/\partial\ln r$. 
In the next section we show that in CGM inflows dominated by turbulence the inflow velocity is constant ($p=0$), and thus $\sigmat=|u_r|/\eta$. In CGM inflows dominated by thermal pressure we show that $p=-0.5$, so we get $\sigmat = |u_r|/(2\eta)$. 

Eqn.~(\ref{eq:sigmat_ur}) also has a solution with $\sigmat\rightarrow0$ and $p\rightarrow -1$. This is the steady-state solution in which the turbulent velocity is small compared to the inflow velocity, so adiabatic heating dominates over dissipation  and $\sigma_t\propto r^{-1}$ (eqn.~\ref{eq:heating}). Such conditions are expected at outer CGM radii before accretion could amplify any initial turbulence field. We thus expect a transition from a flow where   adiabatic heating dominates at outer CGM radii to a flow where adiabatic heating is balanced by dissipation at inner CGM radii. Defining $r^\star$ as the radius where $\sigmat\approx |u_r|$ we get:
 \begin{eqnarray}
 \sigmat\left(r > r^\star\right) &\propto& r^{-1}\nonumber\\
 \sigmat\left(r < r^\star\right) &=& \frac{1+p}{\eta}|u_r| ~.
 \label{eq:sigmat steady state}
 \end{eqnarray} 

\subsection{Non-turbulent CGM inflows}\label{sec:inflows}

In the previous section we argued that in a contracting medium the turbulent velocity roughly equals the inflow velocity (eqn.~\ref{eq:sigmat_ur}). To estimate the inflow velocity we utilize the spherical CGM inflow solutions from \cite{stern19,stern20}, which are based on cooling flow theory developed for the ICM \citep{mathews78,fabian84}. As shown by \cite{stern20}, these inflow solutions change qualitatively at the mass scale of galaxies where cooling times equal dynamical times. We review these non-turbulent inflow solutions here, and  in the next subsection discuss how they are modified by turbulence.

The CGM inflow solutions are derived from the steady-state, spherically-symmetric fluid equations of radiating, non-magnetized ideal gas, with no heat conduction or viscous stress:
\begin{equation}
     \frac{\partial}{\partial r} \left( \rho\ur r^2 \right) = 0
     \label{eq:mass_eq}
\end{equation}
\begin{equation}
     u_r \frac{\partial u_r}{\partial r} + \frac{1}{\rho} \frac{\partial P}{\partial r}  + \frac{{V_\text{c}}^2}{r} = 0
     \label{eq:momentum_eq}
\end{equation}
\begin{equation}
     u_r \frac{\partial 
     \ln
K}{\partial r}  = \frac{1}{t_\mathrm{cool}}
     \label{eq:entropy}
\end{equation}
where $\rho$ is the gas density,  $P=\rho c_{\rm s}^2/\gamma$ is the gas pressure, $\gamma=5/3$ is the adiabatic index, $K=P\rho^{-\gamma}$  is the exponential of the gas entropy up to a constant, $\cs$ is the adiabatic sound speed
\begin{equation}
    c_\text{s}=\sqrt{\frac{\gamma k_\text{B} T}{\mu m_{\rm p}}} ~,
	\label{eq:soundspeed_sim}
\end{equation}
$T$ is the temperature, and
$\mu$ is the molecular weight assumed to equal $0.6$.
The circular velocity $V_{\rm c}$ is 
\begin{equation}
    V_{\rm c}=\sqrt{\frac{GM(<r)}{r}} ~,
\end{equation}
where $G$ is the gravitational constant and $M(<r)$ is the total mass within radius $r$. 
The cooling time $t_{\rm cool}$ is defined as the ratio of the energy per unit volume $(\gamma-1)^{-1}P$ to the radiated energy per unit volume $n_{\rm H}^2\Lambda$:
\begin{equation}
   t_{\mathrm{cool}} = \frac{P}{(\gamma-1) n_\mathrm{H}^2\Lambda} ~,
    \label{eq:cooling_time}
\end{equation}
where $\Lambda$ is the net cooling function, equal to the sum of radiative cooling processes minus radiative heating processes and $n_{\rm H}$ is the hydrogen volume density. For optically thin gas in the CGM heating and cooling rates reach equilibrium at $T_{\rm eq}\approx10^4 \, \mathrm{K}$, which corresponds to:
\begin{equation}
   c_{\rm s,eq} = 15 \left(\frac{T_{\rm eq}}{10^4\,{\rm K}}\right)^{0.5}\, \mathrm{km \, s^{-1}}.
    \label{eq:cs_eq}
\end{equation}
We also define a characteristic free-fall time $t_\mathrm{ff}$
\begin{equation}
    {t}_{\mathrm{ff}}\equiv \frac{\sqrt{2} r}{V_\text{c}} ~,
    \label{eq:t_ff}
\end{equation}
which roughly equals the dynamical time.

Before discussing the solution to eqns.~(\ref{eq:mass_eq}) -- (\ref{eq:entropy}), we review the range of radii and range of halo masses under which they are potentially relevant. 
Equations ~(\ref{eq:mass_eq}) -- (\ref{eq:entropy}) do not account for angular momentum support, and thus are limited to radii larger than the circularization radius $R_{\rm circ}$, defined as the radius where the average tangential velocity $u_{||}$ equals $\vc$:
\begin{equation}\label{eq:Rcirc}
R_{\rm circ} = r(u_{||}=\vc) ~.
\end{equation}
For CGM spin comparable to that of the dark matter halo as seen in cosmological simulations we expect $$
R_\mathrm{circ}\lesssim 0.1\,R_\mathrm{vir}$$ \citep{Stewart13,Stewart17,DeFelippis20,stern20,Stern24}. Second, for radiative losses to be significant and allow an inflow, we limit the discussion to radii where $\tcool$ is smaller than the age of the system ($\sim t_{\rm hubble}$), i.e.~to radii within the cooling radius:
\begin{equation}\label{eq:Rcool}
R_{\rm cool} = r(\tcool=t_{\rm hubble}) ~.
\end{equation}
The value of $R_{\rm cool}$ is typically $0.5\Rvir$ or larger in galaxy-scale halos at $z\sim0$ (see section 2 in \citealt{Stern21A}). We also limit the discussion to halos with $\vc\gg c_{\rm s,eq}$ so virial temperature gas can effectively cool, corresponding to halo masses above a threshold $M_{\rm eq}\approx(1+z)^{-1.5}10^{9}\msun$. Last, the assumption of a quasi-spherical  outer boundary is justifiable when the halo resides in a wide filament with size $\sim\Rvir$ or larger. This is expected in halos which are not high sigma-peak halos, i.e. halos with mass $M_{\rm halo}<M_{\rm grav}$ where $M_{\rm grav}\approx 10^{13}\msun$ at $z\sim0$, $M_{\rm grav}\approx10^{12}\msun$ at $z\sim1$, and $M_{\rm grav}\approx0.5\cdot10^{11}\msun$ at $z\sim2$ (see e.g.\ figure~2 in \citealt{Dekel06}). Above this mass scale halos are penetrated by narrow filaments so the outer boundary deviates substantially from spherical. 

Our analysis also does not account for the dynamical effects of galaxy feedback. The effect of feedback on our results is addressed in the discussion and assessed using the FIRE cosmological zoom simulations in a companion paper \citep{Kakoly25}.

\begin{figure*}
	
	\includegraphics[width=\textwidth]{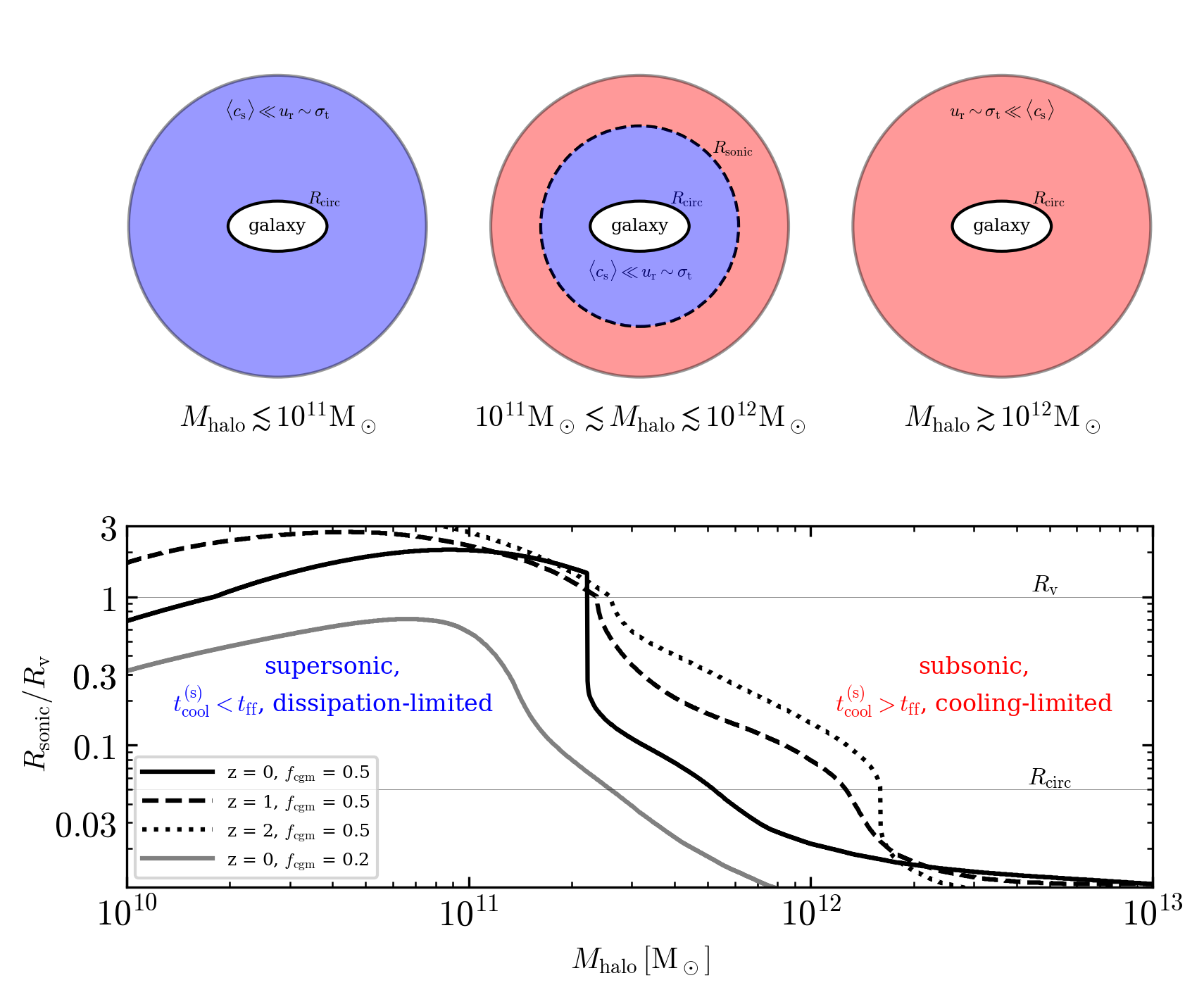}
\vspace{-0.5cm}
\caption{Subsonic and supersonic turbulence in CGM inflows. \textit{Top}: Cartoons of the subsonic and supersonic regimes versus halo mass. At high masses the hot phase ($T\approx T_{\rm vir}$) dominates the inflow, with subsonic inflow and turbulent velocities and an inflow velocity  set by the cooling time (eqn.~\ref{eq:cooling_flow}). At low masses the cool phase ($T\approx T_{\rm eq}$) dominates so inflow and turbulence velocities are supersonic, and the inflow velocity is set by the turbulence dissipation time (eqn.~\ref{eq:ur cool solution}). At intermediate masses CGM inflows have a sonic radius separating the two regimes. 
\textit{Bottom}: Sonic radius of CGM inflows relative to $\Rvir$ versus halo mass and redshift, calculated using eqn.~(\ref{eq:r_sonic}) with $Z=0.3Z_\odot$ and a CGM mass relative to the halo baryon budget of $0.5$ (black lines) or $0.2$ (gray line). Horizontal lines mark $\Rvir$ and the typical circularization radius of CGM inflows, at which the inflow halts and within which the galaxy forms.  
}
    \label{fig:CGM_phases}
\end{figure*}

 \subsubsection{Hot inflows}
 Equations~(\ref{eq:mass_eq}) -- (\ref{eq:entropy}) admit two asymptotic solutions as discussed in \cite{stern20}: a slow cooling solution in which the inflow is hot and contracts quasi-statically, and a solution where cooling is rapid so the inflow is free-falling. We discuss the hot solution here. The cool solution and the transition between the two limits are discussed in the next subsections. 

 The slow cooling limit is achieved when the cooling time of `hot' gas with $\cs^2\approx \vc^2$ (or equivalently $T\approx T_{\rm vir}$) is longer than $\tff$. In this limit the accretion flow remains hot since compressive heating balances radiative cooling, so thermal pressure support is significant and infall is slower than free-fall:
 \begin{equation}
     u_{r,\,{\rm hot}}\approx -\frac{r}{\tcool} = -\frac{\tff}{\sqrt{2}\,\tcool}\vc \sim r^{-0.5}~,
         \label{eq:cooling_flow}
 \end{equation}
where the approximation and index of $-0.5$ are exact for an isothermal potential (constant $\vc$) and a CGM metallicity $Z$ independent of radius. The temperature and density of the gas in the slow cooling limit satisfy 
\changed{
\begin{equation}
    T_{\rm hot}=\frac{4}{3}T_{\rm vir}, \ \ \ \rho_{\rm hot}\propto r^{-1.5}
\label{eq:cooling_flow T and rho}
\end{equation}
where  
\begin{equation}
     \Tvir=\frac{1}{2} \frac{\mu m_\mathrm{p}V_\mathrm{c}^2}{k_\mathrm{B}}~.
\end{equation}}
The implied sound speed satisfies
\begin{equation}
c_{\rm s, hot}^2=\frac{10}{9}\vc^2
\label{eq:cooling flow c_s}
\end{equation}
so the radial mach number is 
\begin{equation}
\mathcal{M}_{r, {\rm hot}}\equiv \frac{\uin}{c_{\rm s}} =0.7\frac{\tff}{\tcool} ~.
\label{eq:mach_r hot}
\end{equation}
This subsonic accretion flow is a form of `hot mode' accretion, and is called a `cooling flow' in  classic ICM literature \citep{mathews78,fabian84}. Note though that despite the name, the flow remains hot and only its entropy decreases with decreasing radius. This type of accretion has been identified in $\Mhalo\gtrsim 10^{12}\msun$ halos at $z\sim0$ in the FIRE zoom simulations \citep{Hafen22,Sultan25}, and extended to a 2D axisymmetric solution which accounts for angular momentum by \cite{Stern24} and \cite{Sankar25}. 

 \subsubsection{Cold inflows}
In the other limit the cooling time of the hot gas is shorter than $\tff$, so compressive heating is not rapid enough to balance radiative cooling. The gas thus cools to the equilibrium temperature of $T_{\rm eq}\approx10^4\,{\rm K}$ and free-falls without pressure support as a `cold flow'\footnote{In some studies, `cold mode' accretion refers to accretion of cold gas directly from the IGM. Here we refer to accretion in this limit as `cold mode' even if the gas is shock-heated and then rapidly cools down.}. For an isothermal potential we thus get
\begin{equation}
    u_{r,\,{\rm cool}}\sim-\frac{r}{\tff}=-\vc,
    \label{eq:cold_flow}
\end{equation}
and
\begin{equation}
    T_{\rm cool}\approx T_{\rm eq},  \ \ \  \rho_{\rm cool}\sim r^{-2} ~.
\label{eq:cold_flow rho and T}
\end{equation}
The inflow is thus supersonic with
\begin{equation}
\mathcal{M}_{r, {\rm cool}} \approx \frac{\vc}{c_{\rm s,eq}}
\label{eq:mach_r cold}
\end{equation}

\newcommand{\fCGM}{f_{\rm cgm}}
\newcommand{\fb}{f_{\rm b}}

\subsubsection{The sonic radius: the transition between the two limits}\label{sec:Rsonic}
The two limits discussed above can form a single flow, with an outer hot flow and an inner cool flow, and the transition occurring at the sonic radius $\Rsonic$ \citep{mathews78,Bertschinger89,stern20}. To estimate $\Rsonic$  we define the cooling time of hot gas $\tcools$ (`s' for shocked hot gas) as the cooling time the gas would have if its temperature was equal to \changed{$(4/3)\,\Tvir$, the expected value in the hot accretion limit (eqn.~\ref{eq:cooling_flow T and rho}):
\begin{equation}
   \tcools \equiv \tcool\left(\frac{4}{3}\Tvir,\nH,Z\right) ~.
    \label{eq:t_cool}
\end{equation}
This definition implies $\tcools=\tcool$ only at radii where the hot accretion limit is valid, i.e.~when $\tcools\gg\tff$.} Now, since tn the hot solution 
the ratio $\tcool/ t_\mathrm{ff}$ decreases inwards as $\sim r^{0.5}$ (see eqns.~\ref{eq:cooling_time}, \ref{eq:t_ff}, and \ref{eq:cooling_flow T and rho}), a flow that at large radii is in the hot limit will transition to the cool limit roughly at the radius
\begin{equation}
\Rsonic= r(t_\mathrm{cool}^{\mathrm{(s)}}= t_\mathrm{ff}) \approx \frac{m_{\rm p}\vc^3}{\nH\Lambda} = 54\,V_{100}^3 n_{-4}^{-1}\Lambda_{-22}^{-1}\,{\rm kpc} ~.
	\label{eq:r_sonic}
\end{equation}
In the approximation in eqn.~(\ref{eq:r_sonic}) we used eqns.~(\ref{eq:t_cool}),  (\ref{eq:cooling_time}), (\ref{eq:t_ff}), and $\rho\approx\sqrt{2}\nH m_{\rm p}$. For calculating the numerical value we defined $V_{100}\equiv \vc/100\kms$, $n_{-4}\equiv\nH/10^{-4}\,{\rm cm}^{-3}$ and $\Lambda_{-22}\equiv\Lambda/10^{-22}\,{\rm erg}\,{\rm s}^{-1}\,{\rm cm}^3$. 

The value of $\Rsonic$ is plotted versus halo mass in Figure~\ref{fig:CGM_phases}, for several redshifts and CGM mass fractions $\fCGM$, where $\fCGM f_{\rm b}M_{\rm halo}$ is the total CGM mass and $\fb=0.16$ is the cosmic baryon fraction. To derive $\Rsonic$ we use eqn.~(\ref{eq:r_sonic}) assuming  $\nH\propto r^{-1.5}$ (eqn.~\ref{eq:cooling_flow T and rho}), while $\vc(r)$ is derived assuming 
an NFW potential for the halo with concentration parameter from \cite{DuttonMaccio14}, plus a galaxy with mass corresponding to the stellar mass -- halo mass relation from \cite{Behroozi19}. The galaxy mass is assumed to be an exponential disc with scale length $0.013\Rvir$ \citep{kravtsov13} and cut off at four scale lengths.  For the calculation of $\Lambda$ we use the \cite{Wiersma09} tables with $T=(4/3)\Tvir$ and a gas metallicity $Z$. \changed{As fiducial values we use $\fCGM=0.5$, comparable to values found in FIRE simulations of Milky-Way mass halos \citep{hafen19}, and $Z=0.3\,{\rm Z}_\odot$ based on estimates of the Milky-Way hot CGM metallicity \citep{Bregman18}, though we emphasize that these values are uncertain. }

Fig.~\ref{fig:CGM_phases} shows that the sonic radius is smaller than $\Rcirc$ at $M_{\rm halo}\gtrsim 10^{12}\msun$ for $Z=0.3\,Z_\odot$ and $\fCGM=0.5$, indicating that the CGM is hot and quasi-static down to the galaxy scale at the Milky-Way mass scale and above. This mass threshold decreases with decreasing $\fCGM$ due to the longer cooling times at lower density. At $M_{\rm halo}\lesssim10^{11}\msun$ and $\fCGM=0.5$ we find $\Rsonic>\Rvir$, indicating that the solution is in the cool limit throughout the CGM. At intermediate masses of $10^{11}-10^{12}\msun$ when $\fCGM=0.5$, or at all masses $<10^{11.5}\msun$ when $\fCGM=0.2$, the solution is `transonic' -- hot and subsonic at outer CGM radii, and cool and supersonic at inner CGM radii.  These three regimes are depicted in the top panels of Fig.~\ref{fig:CGM_phases} and were discussed previously in \cite{stern20}. These three regimes were also 
identified in the FIRE cosmological zoom simulations by \cite{Stern21A,Stern21B}. \changed{The main addition of the current paper to this picture is that at radii smaller than $\Rsonic$, the cool inflow tends to be turbulence-dominated rather than free-falling, as discussed in the following subsection.}

\subsection{Turbulent CGM inflows}\label{sec: turb inflows}
We now combine the evolution of turbulence in contracting flows discussed in section~\ref{sec:turb} with the two inflow solutions discussed in section~\ref{sec:inflows}. Our goal is to derive the radial profile of the turbulent velocity and its effect on CGM inflows. 

\subsubsection{Hot turbulent inflows}\label{sec:hot}

The hot accretion limit has $u_r^2\ll c_{\rm s}^2$ (eqn.~\ref{eq:cooling_flow}), so inertial terms are small compared to thermal pressure in the momentum equation (eqn.~\ref{eq:momentum_eq}), and kinetic energy is a small part of the flow energy budget. The expectation that $\sigma_t\sim|u_r|$ thus implies that turbulence pressure and energy are also small relative to thermal pressure. We thus expect $T(r)$, $\rho(r)$ and $u_r(r)$ to be similar to those in the non-turbulent solution in eqns.~(\ref{eq:cooling_flow}) -- (\ref{eq:cooling_flow T and rho}). The turbulent velocity can then be derived from this non-turbulent solution and eqn~(\ref{eq:sigmat steady state}):
\begin{equation}
   \sigma_\text{t,\,{\rm hot}}\approx-\frac{1}{2\eta}u_\mathrm{r}\approx-\frac{r}{2\eta\tcool}\propto r^{-0.5} ~.
    \label{eq:cooling_flow_turb}
\end{equation}
Using eqn.~(\ref{eq:cooling flow c_s}) we get that the turbulence Mach parameter is equal to:
\begin{equation}
    \mathcal{M}_\mathrm{t,\,{\rm hot}} = \frac{9}{10}\frac{\sigma_{\rm t,\,hot}}{\vc}= 0.3\,\eta^{-1}\frac{\tff}{\tcool} ~.
\label{eq:mach_t hot}
\end{equation}

\subsubsection{Cool turbulent inflows}\label{sec:cool}

In the rapid cooling limit where $\tcools\ll \tff$ the flow is cool with $u_r^2\sim \vc^2 \gg c^2_{\rm s}$ (eqn.~\ref{eq:cold_flow}), indicating that thermal pressure is negligible relative to inertial forces. The expectation that $\sigma_t\sim|u_r|$ thus indicates that turbulent pressure is substantial, and we need to modify  the momentum equation to account for turbulence. For isotropic turbulence with negligible thermal pressure the momentum equation is 
\citep[e.g.,][]{Hennebelle21,Sultan25}
\begin{equation}
     \ur \frac{\partial \ur}{\partial r} + \frac{1}{\rho} \frac{\partial}{\partial r} \left( \frac{1}{3}\rho {\sigma_\text{t}}^2 \right) + \frac{{V_\text{c}}^2}{r} = 0~,
\label{eq:momentum_eq turb}
\end{equation}
where the factor of $1/3$ in the second term originates from our 3D definition of $\sigmat$. Note that for anisotropic turbulence, the $\sigmat^2/3$ term should be replaced with the 1D radial component of the turbulence $\sigma_{{\rm t},r}$ and another term equal to $2\sigma_{{\rm t},r}^2-\sigma_{{\rm t},||}^2$ should be added, where $\sigma_{t,||}$ is the tangential component of the turbulence (\citealt{Hennebelle21}, see also \citealt{Lau13}). We  disregard this complication here since in our simulations  significant anisotropy appears only when the total support from turbulence is small, either in the hot inflow regime or at large CGM radii where the flow is in free-fall (see below).

When $\sigma_t^2\ll u_r^2$ the turbulent term in eqn.~(\ref{eq:momentum_eq turb}) is negligible compared to the inertial term, and the inflow velocity equals the free-fall velocity as discussed above. The turbulent velocity then follows
\begin{equation}
   \sigma_\text{t,\,{\rm ff}}  \propto r^{-1}
    \label{eq:sigmat free-fall}
\end{equation}
as explained in section~\ref{sec:turb}, where `ff' emphasizes that this relation is expected in free-fall. 
This rapid increase in $\sigmat$ implies that at small enough radius $\sigmat$ will reach  $\sim \uin$, so turbulence pressure becomes significant and the relation between $\sigmat$ and $\ur$ in eqn.~(\ref{eq:sigmat_ur}) will hold. Using this relation in the momentum conservation equation (\ref{eq:momentum_eq turb}), together with mass conservation (eqn.~\ref{eq:mass_eq}) and an isothermal potential yields
\begin{equation}
      u_{r,\,{\rm td}}\approx -\sqrt\frac{3}{2}\eta V_\text{c}  , \quad \sigma_\text{t,\,td}\approx \sqrt\frac{3}{2}V_\text{c}, \quad \rho_{\rm td}\propto r^{-2},
     \label{eq:cool_solution}
\end{equation}
where `td' stands for `turbulence-dominated'. In this solution the turbulence pressure gradient entirely balances the gravitational acceleration so the inflow is not accelerating.
Dividing by $c_{\rm s,eq}$ then gives the mach numbers of the solution
 \begin{equation}
    \mathcal{M}_{r,\,{\rm td}}=\sqrt\frac{3}{2}\eta \frac{V_\mathrm{c}}{ c_\mathrm{s,\, eq}},\quad  \mathcal{M}_\mathrm{t,\,td}=\sqrt\frac{3}{2}\frac{V_\mathrm{c}}{ c_\mathrm{s,\, eq}}.
     \label{eq:cool_machs}
\end{equation}

Eqn.~(\ref{eq:cool_solution}), combined with the definition of the dissipation time (eqn.~\ref{eq:tdiss}), implies
\begin{equation}\label{eq:ur cool solution}
    u_{r,\,{\rm td}} = -\frac{r}{\tdiss} ~,
\end{equation}
which demonstrates that the inflow velocity in this solution is set by the turbulence dissipation time, in contrast with being set by the cooling time as found in the hot limit (eqn.~\ref{eq:cooling_flow}). 
We thus conclude that in the inner CGM of $<10^{12}\msun$ halos where cooling is rapid, the inflow is not free-falling or thermal pressure supported,  but rather partially supported by turbulence with $\sigmat\approx\vc$.  
This $\sigmat\approx |u_r|\approx\vc$ result in the cool inflow regime is noted in the top panels of Fig.~\ref{fig:CGM_phases} and is a main conclusion of this paper.

\subsection{Density and temperature distributions}\label{sec:density_distribution}

Our result above that inner CGM of $\lesssim10^{12}\msun$ halos are highly turbulent imply a large density dispersion at a given radius, on top of any density dispersion induced by cooling physics and pressure equilibrium arguments. In this section we estimate this turbulence-induced density dispersion, based on the expectations of isothermal turbulence. The  isothermal approximation is reasonable at $\lesssim10^{12}\msun$ due to the rapid cooling as discussed above, and as further demonstrated below using the radiative simulations. 
We do not consider self-gravity effects which are expected to be small at least in the low-redshift CGM \citep{Stern16,Chen23}. Self-gravity may be more important at $z>4$ where densities are higher \citep{Sun25}. 

In simulations of isothermal and homogeneous turbulence where turbulence is explicitly driven at large scales, the mass-weighted density distribution can be approximated with a lognormal distribution \citep[e.g.,][]{Krumholz2014}:
\begin{equation}
\mathrm{PDF}(s)=\frac{1}{\sqrt{2\pi} \sigma_\mathrm{s}}\mathrm{exp}\left[ -\frac{(s-s_0)^2}{2\sigma_\mathrm{s}^2}\right]
	\label{eq:PDF}
\end{equation}
where $s\equiv\mathrm{ln}(\rho/\rho_0)$, $\sigma_\mathrm{s}$ is the standard deviation of $s$, $\rho_0$ is the volume-weighted mean density and $s_0=-\sigma_\mathrm{s}^2/2$. The standard deviation is related to the turbulent mach number by:
\begin{equation}
\sigma_\mathrm{s}=\sqrt{\mathrm{ln}\left(1+b^2\mathcal{M}_\mathrm{t}^2\right)} ~.
	\label{eq:sigma_log_rho}
\end{equation}
The constant $b$ is found to depend on the nature of the external stirring force of the turbulence $\vec{f}_\mathrm{stir}$ in the simulation, 
with $b=1/3$ for purely solenoidal driving ($\vec{\nabla}\cdot \vec{f}_\mathrm{stir}=0$) and $b=1$ for purely compressive driving ($\vec{\nabla}\times \vec{f}_\mathrm{stir}=0$, \citealt{Federrath08}).
In accretion-driven turbulence the only external driving force is gravity which is compressive, \changed{ so one may expect $b \approx  1$. The simulations below suggest somewhat lower values of $0.5-0.8$. }

Using eqn.~(\ref{eq:sigma_log_rho}) with the estimate of $\mathcal{M}_{\rm t}$ in the turbulence-dominated regime (eqn.~\ref{eq:cool_machs}) we get:
\begin{equation}
\sigma_\mathrm{s, cool}=\sqrt{\mathrm{ln}\left(1+\frac{3}{2}\left(\frac{\changed{b}V_\mathrm{c}}{c_\mathrm{s,eq}}\right)^2\right)}\approx 2.2\left(\frac{\changed{b}V_\mathrm{c}/c_{\mathrm{s,eq}}}{10}\right)^{0.25}.
	\label{eq:sigma_log_rho_cold}
\end{equation}
where the approximation is accurate to 5\% for $3<\changed{b}V_\mathrm{c}/c_\mathrm{s,eq}<20$. 
We thus expect a wide density distribution with $\sigma_\mathrm{s}\approx 2$ in the turbulence-dominated regime, or equivalently a full width half max (FWHM) of about two decades in density, at any given CGM radius. This density dispersion is comparable to that expected due to mixing between CGM phases \citep[see, e.g.][]{Faerman25}. 

In the hot inflow regime Mach numbers are subsonic, so turbulence-driven density fluctuations are expected to be small. Using eqn.~(\ref{eq:mach_t hot}) in eqn.~(\ref{eq:sigma_log_rho}) we get
\begin{equation}
\sigma_\mathrm{s, hot}\approx \changed{b}\mathcal{M}_\mathrm{t}\approx\frac{\changed{b}}{2\eta}\frac{t_\mathrm{ff}}{t_\mathrm{cool}}~.
	\label{eq:sigma_log_rho_hot}
\end{equation}
For the $\eta=0.2$ found below in the subsonic regime, \changed{and assuming $b=0.5$ and a typical $\tcool/\tff = 10$ we get $\sigma_\mathrm{s, hot}\approx0.1$.}

Regarding the temperature distribution, we note that in non-turbulent inflows the gas temperature is unimodal at any given radius: the temperature of the inflow switches abruptly from $T\approx T_{\rm vir}$ outside the sonic radius to $T\approx T_{\rm eq}$ within the sonic radius \citep{stern20}. 
Turbulence in the inflow is expected to widen the temperature distribution since the wide density distribution implies a wide range of cooling times  at each radius, and since turbulence dissipation adds another source of heat to the gas.  Turbulent inflows are thus expected to be multi-phase as seen in the CGM in galaxy formation simulations, with the dominant phase set by $\tcools/\tff$ at mean density.

\subsection{Velocity Structure Functions}\label{sec:VSF analytic}

\newcommand{\du}{\Delta u}
\newcommand{\dr}{\ell}

The first-order velocity structure function (VSF) is defined  as the statistical relation between the difference in velocity $\du=|\vec u_1-\vec u_2|$ and the physical separation $\dr=|\vec{r}_2-\vec{r}_1|$ of pairs of fluid elements $(1,2)$. On scales below the driving scale and above the viscous scale (the `inertial' range), the VSF is expected to have a Kolmogorov spectrum ($\du\propto \dr^{1/3}$) in subsonic turbulence which is isotropic and homogeneous, while a Burgers spectrum is expected ($\du\propto\dr^{1/2}$) for supersonic turbulence which is isotropic and homogeneous. What are the driving scale and inertial range in accretion-driven turbulence, where turbulence is enhanced by adiabatic heating? 

The adiabatic heating term (eqn.\ \ref{eq:heating}) is scale-independent, so in principle a contracting medium
does not have an `inertial range' -- at all scales some of the energy is a result of adiabatic heating in addition to the cascade from larger scales. 
However, a derivation of the VSF that accounts for both the cascade and adiabatic heating in the turbulence-dominated solution yields (see Appendix~\ref{app:VSF})
\changed{\begin{equation}\label{eq: VSF with heating}
\du(\dr,r) = \sigmat\left[\frac{5}{3}\left(\frac{\ell}{r}\right)^{1/2} - \frac{\ell}{r}\right] ~,
\end{equation}
where $\du(\dr,r)$ is the VSF measured at radius $r$.  This VSF is flatter than the Burgers spectrum at scales $\dr\sim r$, but converges to the Burgers scaling  at smaller scales of $\dr\ll r$. The effect of adiabatic heating is thus small at scales $\dr\ll r$ and the cascade from larger scales dominates, so these scales are effectively inertial as long as they are larger than the viscous scale. The `driving scale' in a contracting medium is thus effectively $\sim r$.}

\section{Hydrodynamic Simulations -- Setup}\label{sec:setup}

In this section, we present a series of 3D hydrodynamic simulations of turbulent, radiatively cooling gas accreting in a dark matter halo potential. Our main goal is to test the analytic results of the previous section: that 
the turbulent velocity of the gas asymptotically approaches the inflow velocity; and that in massive halos 
the turbulence is subsonic while in low mass halos the turbulence is supersonic.

\subsection{Code and Physics}
The simulations are performed using the multi-method gravity and hydrodynamics code GIZMO \citep{hopkins15} in its meshless finite-mass mode (MFM). MFM is a Lagrangian, mesh-free, finite-mass method that combines advantages of traditional smooth particle hydrodynamics (SPH) and grid-based methods.
The physics of the simulations are set as follows: To mimic the dark-matter halo and galaxy potential, we use a fixed gravitational acceleration corresponding to an infinite isothermal sphere, which gives rise to a radius-independent $V_\mathrm{c}$ profile.
For the cooling function $\Lambda$ we use the tables from \cite{Wiersma09}, which include detailed heating and cooling rates for 11 independent species (H, He, C, N, O, Ne, Mg, Si, S, Ca, Fe) assuming optically thin conditions and the UV background from \citet{FaucherGiguere09}. This gives $\Lambda$ as a function of $z$, $T$, $\rho$ and $Z$.
We introduce velocity perturbations in the initial conditions to seed the turbulence as described below, though we emphasize there is no turbulence driving during the simulations. Rather, the inflow maintains and `heats' these perturbations into a turbulent velocity field, and is therefore the only source that maintains the turbulence in these simulations. We do not include gas self-gravity, magnetic fields, thermal conduction, explicit viscosity, or feedback from the galaxy, leaving exploration of the effects of these processes to future work (see discussion).

\subsection{Initial Conditions}\label{sec:ICs}

We run simulations of gas in halos with masses in the range $10^{10}\, \mathrm{M_\odot}\le M_\mathrm{halo}\le10^{13}\,\mathrm{M_{\odot}}$ at redshifts $0\le z \le 2$, listed in Table~\ref{tab:sims_table}. From 
$M_{\rm halo}$ and $z$ the virial radius $\Rvir$ and circular velocity $\vc$ are computed using the relations outlined in Appendix A2 and section 3.1 of \citet{Dekel06}.  We emphasize that these are not cosmological simulations and that the redshift is used solely to determine $\vc$ and $\Rvir$, while these parameters stay constant throughout the simulation. 

Each simulation employs $2\cdot10^6$ 
resolution elements of equal baryonic mass $m_{\rm b}$. They are initially distributed in radial logarithmic shells from $0.015R_\mathrm{v}$ to $10R_\mathrm{v}$, with a random solid angle for each resolution element and an initial density profile set to
\begin{equation}\label{eq:IC density}
\rho_0 =\frac{\fCGM\fb M_\text{halo}}{4\pi R_\text{v}} r^{-2} ~,
\end{equation}
where $\fb$ is the cosmic baryon mass fraction and $\fCGM=0.5$. 
The initial sound speed $c_{\rm s,0}$ and pressure $P_0$ are set by the hydrostatic relation
\begin{equation}
    \frac{\partial\, \text{ln}\, P_0}{\partial\, \text{ln}\, r}=-\gamma \frac{V_\text{c}^2}{c_\text{s,0}^2}.
	\label{eq:hydrostatic}
\end{equation}
For the density profile in eqn.~(\ref{eq:IC density}) and a constant sound-speed, we get $c_{\mathrm{s},0} = \sqrt{5/6} \, \vc$.
The metallicity is set to $0.3Z_\odot$, with relative metal abundances equal to solar values.
\changed{Our choice of $Z$ and $\fCGM$ is discussed in section~\ref{sec:Rsonic} and mainly affects the location of the sonic radius at a given halo mass (Fig.~\ref{fig:CGM_phases}). }

The velocity field is initialized with a lognormal spectrum in k-space: 
\begin{equation}
u_\mathrm{i}(\vec{k})\propto \mathrm{WGN}(\vec{k})\times\frac{k^{-1} }{\sqrt{2\pi \sigma_{\log  k}^2}}\mathrm{exp}\left[ -\frac{\log\left(k/k_\mathrm{peak}\right)^2}{2\sigma_{\log k}^2}\right]
	\label{eq:pertubation_spectrum}
\end{equation}
where $\log$ is shorthand for $\log_{10}$, $\mathrm{WGN}(\vec{k})$ is white Gaussian noise generated as a normal distribution of amplitudes and uniform distribution of phases for each $\vec{k}$, $\sigma_{\log k}$ is the standard deviation of the distribution and $k_\mathrm{peak}$ is the mode of the distribution. We choose a narrow $\sigma_{\log k}=\log2$ in order to have a defined scale for the initial perturbations. The velocity perturbation field is normalized to have an RMS velocity of the desired initial turbulent velocity $\sigmat^0$.
Most of the simulations are set with $\sigma_{{\rm t},0}=0.2\vc$, which is substantially lower than the expected steady state value of $\sqrt{3/2}\vc$ in the cold mode (eqn.~\ref{eq:cool_solution}).  
This corresponds to initial turbulent velocities of $\sim10-50\,  \mathrm{km \, s^{-1}}$. 
Most of the simulations are set with an initial turbulence scale of $\ell_{\mathrm{t,0}} =2\pi /k_\mathrm{peak}= 3\Rvir$. We run also simulations with other $l_{{\rm t},0}$ and $\sigma_{{\rm t},0}$ to explore how they affect our results (see Table~\ref{tab:sims_table}). 
The initial velocity perturbations are set to have a solenoidal fraction $f_{\mathrm{solenoidal}}$ by dividing to the solenoidal and compressive terms and summing them with weight $f_{\mathrm{solenoidal}}$ for the solenoidal perturbations and $1-f_{\mathrm{solenoidal}}$ for compressive perturbations. Since solenoidal component has two degrees of freedom and compressive component has one, the initial solenoidal fraction is set to $f_{\mathrm{solenoidal}} = 2/3$, ensuring equal energy per degree of freedom between solenoidal and compressive components. Testing suggests that our results are insensitive to $f_{\mathrm{solenoidal}}$ in the initial conditions, likely since after an initial transition period turbulence the only external force is gravity which is fully compressive.

\setlength{\tabcolsep}{4pt}  
\changed{
\begin{table}
\footnotesize
    \centering
    \begin{tabular}{l@{}ccccccc@{}} 
        \hline
        (1) & (2) & (3) & (4) & (5) & (6) & (7) & (8)\\
        Name & $\log M_\text{halo}$ & $z$ & $\vc$ & $R_{\rm v}$ & $\sigma_\text{t0}$ & $\ell_\text{t0}$ & $m_{\rm b}$ \\
         & $[\mathrm{M}_\odot]$ & & $[\rm{km}\,{\rm s}^{-1}]$ & $[\mathrm{kpc}]$ & $[\vc]$ & $[R_{\rm v}]$ & $[10^4\,{\rm M}_\odot]$\\
        \hline
        m10z0   & 10   & 0 & 23  & 78  & 0.2  & 3 & 0.38 \\
        m10z1   & 10   & 1 & 36  & 33  & 0.2  & 3 & 0.38 \\
        m10z2   & 10   & 2 & 45  & 22  & 0.2  & 3 & 0.38 \\
        m105z0  & 10.5 & 0 & 34  & 115 & 0.2  & 3 & 1.2  \\
        m105z1  & 10.5 & 1 & 53  & 49  & 0.2  & 3 & 1.2  \\
        m105z2  & 10.5 & 2 & 66  & 32  & 0.2  & 3 & 1.2  \\
        m11z0   & 11   & 0 & 50  & 169 & 0.2  & 3 & 3.8  \\
        m11z1   & 11   & 1 & 77  & 72  & 0.2  & 3 & 3.8  \\
        m11z2   & 11   & 2 & 96  & 46  & 0.2  & 3 & 3.8  \\
        m115z0  & 11.5 & 0 & 74  & 248 & 0.2  & 3 & 12   \\
        m115z1  & 11.5 & 1 & 110 & 105 & 0.2  & 3 & 12   \\
        m115z1\_a$^\star$ & 11.5 & 1 & 110 & 105 & 0.2  & 3 & 12 \\
        m115z1\_b & 11.5 & 1 & 110 & 105 & 0.08 & 3 & 12   \\
        m115z1\_c & 11.5 & 1 & 110 & 105 & 0.2  & 1 & 12   \\
        m115z1\_d & 11.5 & 1 & 110 & 105 & 0.2  & 3 & 1.5  \\
        m115z1\_e & 11.5 & 1 & 110 & 105 & 0.2  & 3 & 97  \\
        m115z2  & 11.5 & 2 & 140 & 68  & 0.2  & 3 & 12   \\
        m12z0   & 12   & 0 & 110 & 364 & 0.2  & 3 & 38   \\
        m12z1   & 12   & 1 & 170 & 155 & 0.2  & 3 & 38   \\
        m12z2   & 12   & 2 & 210 & 100 & 0.2  & 3 & 38   \\
        m125z0  & 12.5 & 0 & 160 & 535 & 0.2  & 3 & 120  \\
        m125z1  & 12.5 & 1 & 250 & 227 & 0.2  & 3 & 120  \\
        m125z2  & 12.5 & 2 & 300 & 147 & 0.2  & 3 & 120  \\
        m13z0   & 13   & 0 & 230 & 785 & 0.2  & 3 & 380  \\
        m13z1   & 13   & 1 & 360 & 333 & 0.2  & 3 & 380  \\
        m13z2   & 13   & 2 & 450 & 216 & 0.2  & 3 & 380  \\
        \hline
    \end{tabular}
\caption{Parameters of the hydrodynamic GIZMO simulations used in this work.
(1) Simulation name; (2) Halo mass; (3) Halo redshift; (4) Circular velocity of isothermal potential; (5) Virial radius; (6) Turbulent velocity in initial conditions in units of $\vc$; (7) Turbulence scale in initial conditions in units of $\Rvir$; (8) Mass of gas resolution elements. \changed{Simulation `m115z1\_a' is isothermal, all other simulations include radiative cooling.}}
\label{tab:sims_table}
\end{table}
}

\begin{figure*}
    \centering
    
    \includegraphics[width=\textwidth]{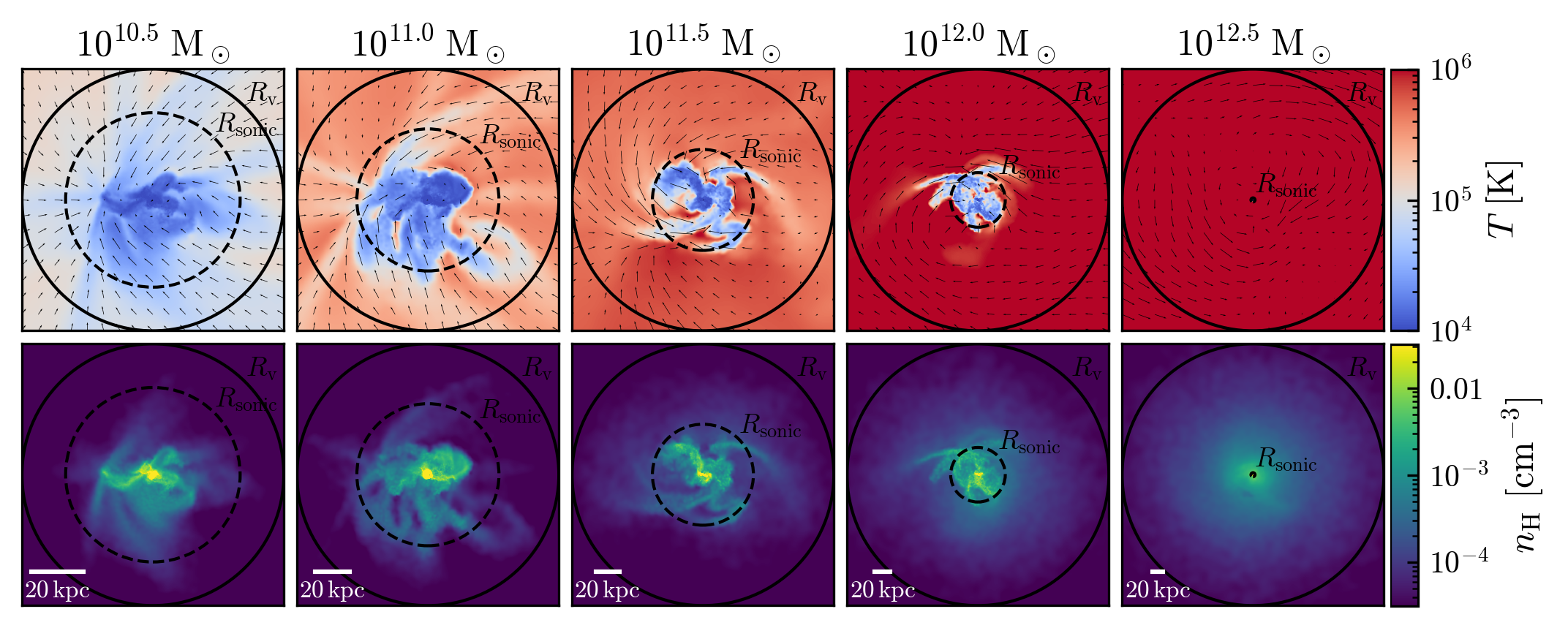}
    
    \caption{Idealized GIZMO simulations of turbulent accretion flows in $z=1$ halos with different masses. Panels show temperature (top) and density (bottom) in snapshots after the flow becomes quasi-steady. 
    Black circles mark the virial radius (solid) and the inflow sonic radius (dashed). Arrows in the top panels mark projected velocity vectors. 
At $r<R_{\rm sonic}$ the CGM is predominantly cool by mass and exhibits large density fluctuations at a given radius. \changed{A movie of the $10^{11.5}\msun$ simulation is available in the supplementary material.}
}
    \label{fig:sim_img}
\end{figure*}

To avoid forming a disk outside our sink radius (see next subsection), the mean angular momentum induced by the initial turbulent velocity field is removed. This is done by subtracting the mean tangent velocity in radial logarithmic shells with relative width of $\delta r/r=0.24$. Exploration of CGM turbulence near the disk where angular momentum is substantial is left for future work.

\begin{figure}
\includegraphics[width=\columnwidth]{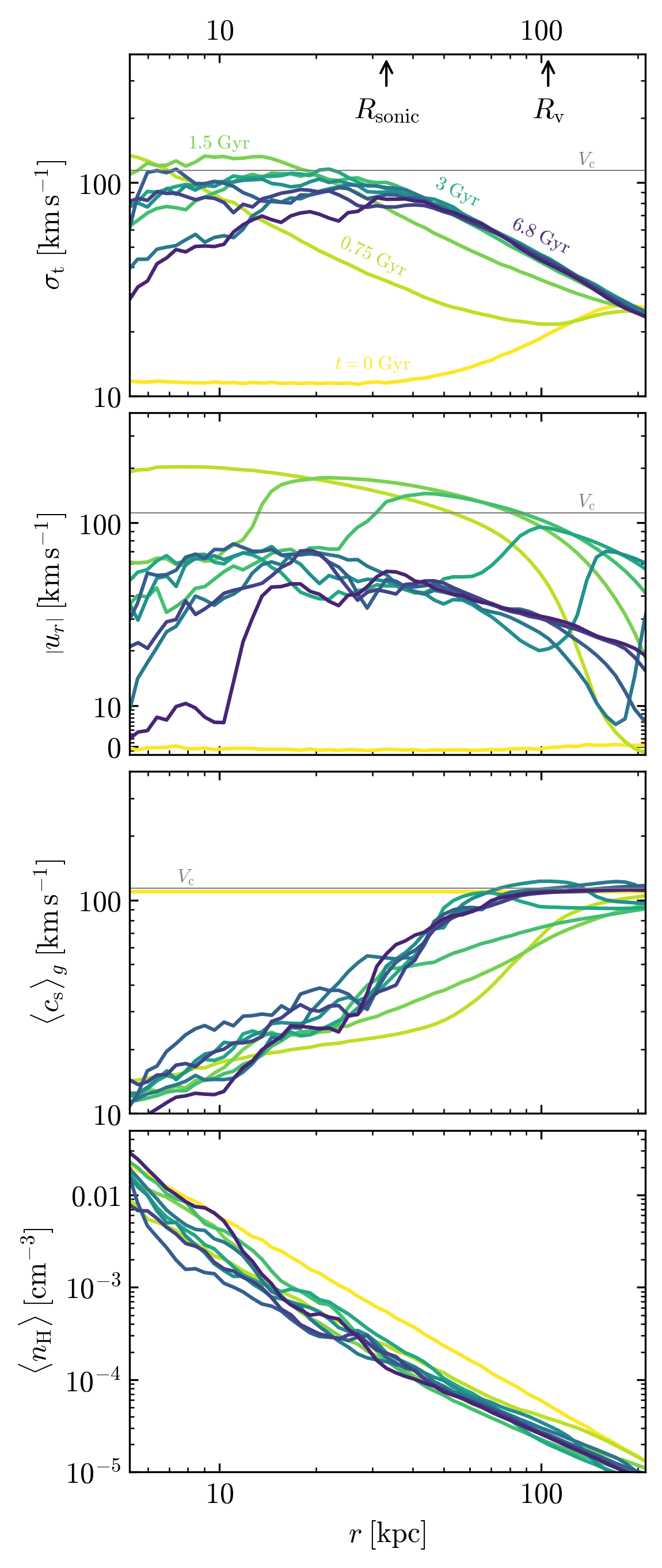}
\vspace{-0.5cm}
\caption{Formation of a turbulent CGM accretion flow in the $10^{11.5}\,\mathrm M_{\odot}$,  $z=1$ simulation.
Panels show \changed{from top to bottom} profiles of the turbulent velocity, average inflow velocity, geometric average sound speed, \changed{and mean hydrogen density} at different snapshots separated by $750\,{\rm Myr}$. Initial conditions (yellow lines) are hydrostatic with a low turbulent velocity of $\approx20\kms$ at large radii $\gtrsim 100\,\rm kpc$. 
Within $\sim2\,{\rm Gyr}$ the turbulent velocity increases to $\sigmat\approx 100\kms\approx \vc$ at small radii of $r\lesssim 30\,{\rm kpc}\approx R_\mathrm{sonic}$. At these inner radii the flow initially cools and free-falls with $\uin\approx 150-200\kms$, and then slows down to $\uin\sim50\kms$ as turbulence develops. After $\approx3\,\mathrm{Gyr}$ all profiles remain constant with time indicating a quasi-steady flow.
}
    \label{fig:sigma_t_time}
\end{figure}

\begin{figure}
\includegraphics[width=\columnwidth]{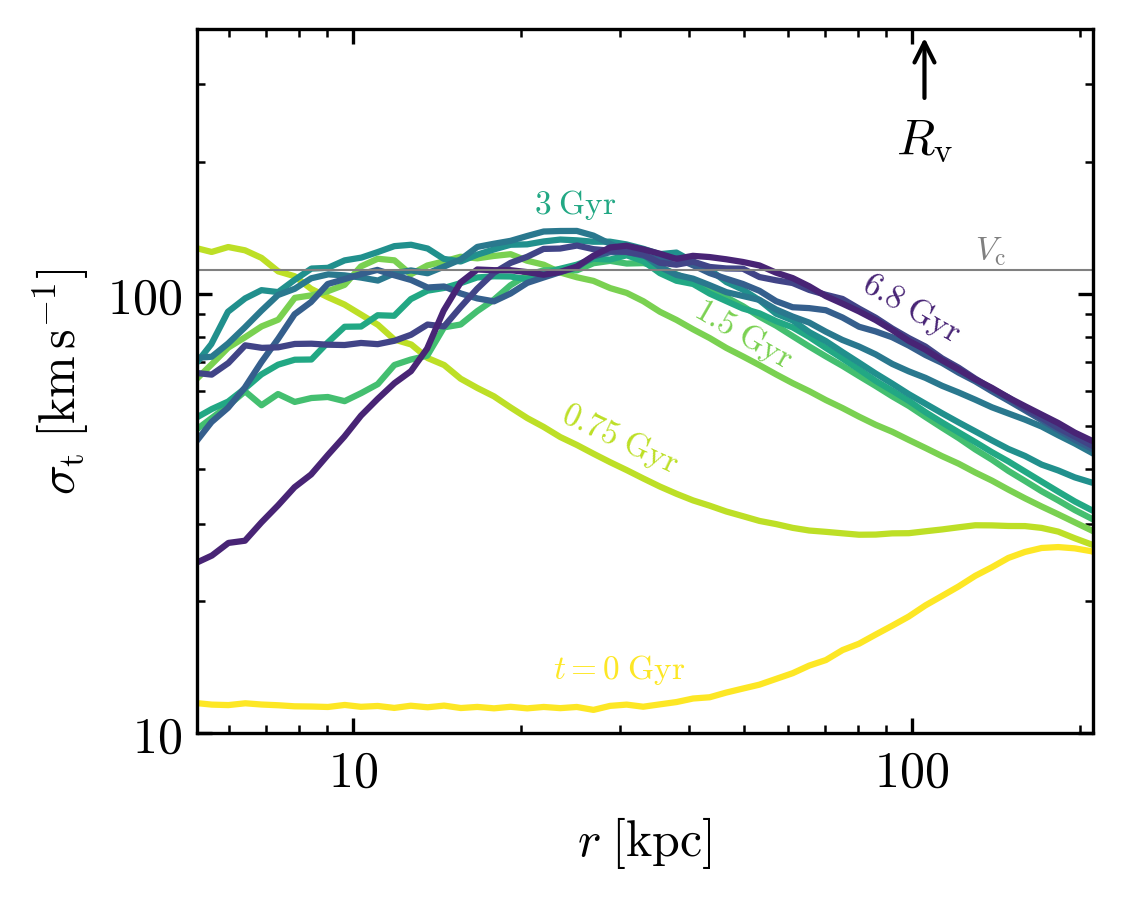}
\vspace{-0.5cm}
\caption{\changed{Formation of a turbulent CGM accretion flow in an isothermal simulation. Simulation parameters and initial conditions are the same as in the radiative simulation shown in Fig.~\ref{fig:sigma_t_time}, except for a constant gas temperature equal to $T_{\rm eq}=10^4\,{\rm K}$. The turbulent velocity profile increases with time similar to the radiative simulation, reaching $\sigmat\approx 100\kms\approx \vc$ at inner radii of $r\lesssim 60\,{\rm kpc}$. This indicates that turbulence in our radiative simulations is driven by adiabatic heating in the inflow, rather than by thermal effects. }
}
    \label{fig:sigma_t_time_iso}
\end{figure}

\begin{figure*}
	
 \centering
	\includegraphics[width=\textwidth]{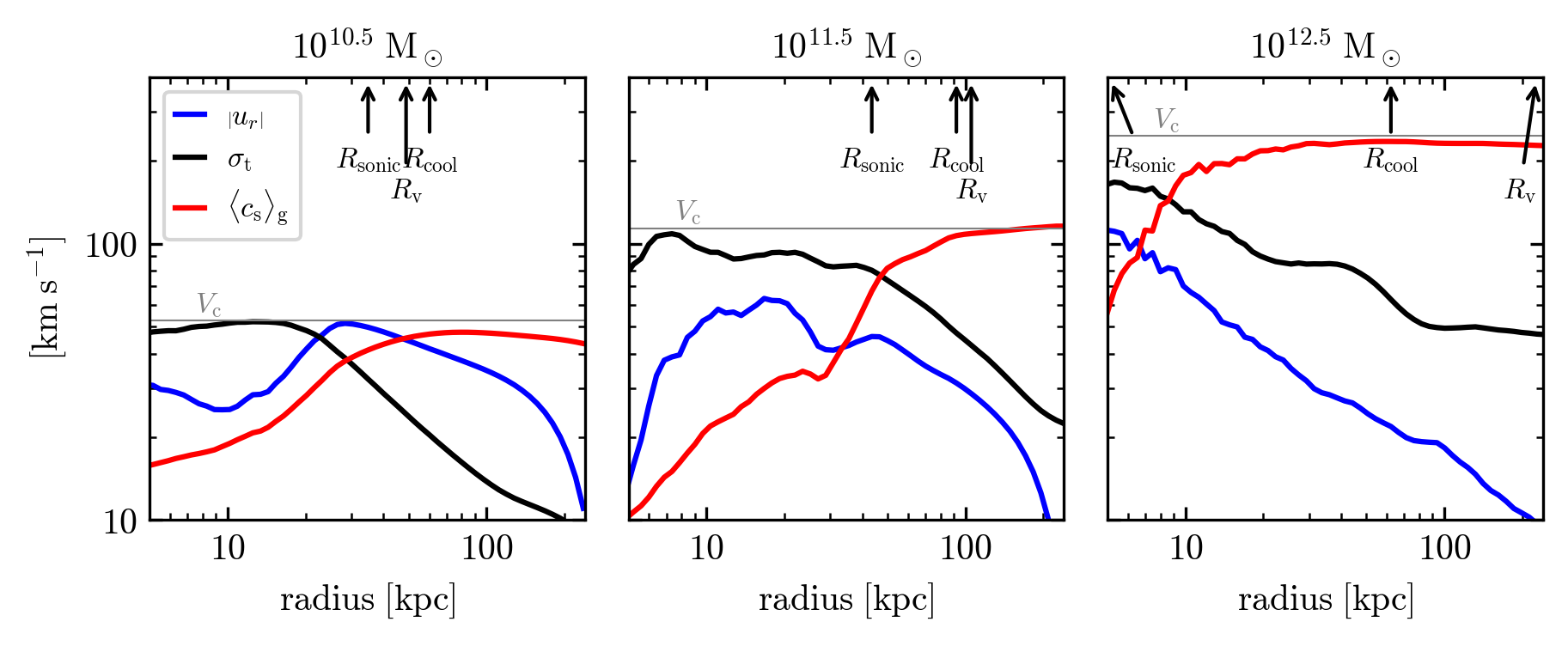}
    \vspace{-0.5cm}
    \caption{Radial profiles of characteristic velocities in turbulent CGM accretion. Panels show turbulent velocity (black), average inflow velocity (blue), and geometric average sound-speed (red) for three $z=1$ simulations with $M_{\text{halo}}= 10^{10.5}\msun$ (left), $10^{11.5}\msun$ (middle), and $10^{12.5}\msun$ (right). Profiles are from snapshots after the flow becomes quasi-steady. Horizontal lines indicate the circular velocity. The sonic radius is on CGM scales in the $10^{10.5}$ and $10^{11.5}\msun$ simulations. Within $\Rsonic$ the inflows have $\sigmat\approx\vc$, cool gas temperatures and constant inflow speeds, implying that the inflow is dominated by turbulence. Note that $\sigmat$ follows $\ur$ up to an order-unity factor in the $10^{11.5}\msun$ simulation, and at small radii in the $10^{12.5}\msun$ and $10^{10.5}\msun$ simulations. }
    \label{fig:velocity_radial_profiles}
\end{figure*}

\subsection{Boundary Conditions}
To mimic the removal of gas at the center of the halo by star formation, we convert gas resolution elements into star particles when they fall within a `sink' radius of $r_{\text{sink}} = 1 \, \text{kpc}$. Within this radius, resolution elements that exceed a number density threshold of $n_{\text{SF}}=0.8 \, \mathrm{cm^{-3}}$ are converted to stars. The density threshold ensures the presence of some gas resolution elements within $r < r_{\text{sink}}$ in order to avoid numerical issues that can arise close to the sink radius. 

The outer boundary of the simulation is treated as an open boundary condition. Resolution elements are initially generated up to a radius of $10R_{\mathrm v}$, to allow continuous accretion of gas onto the inner $R_{\mathrm v}$ sphere for a time of at least $\sim 10t_\mathrm{ff}(R_\mathrm{v})$. 
The simulation analysis is limited to the region $0.05-2\,R_{\mathrm v}$, which avoids the effects of the outer boundary on the results.

\section{Hydrodynamic Simulations -- Results}\label{sec:results}

Figure~\ref{fig:sim_img} shows the results of five simulations of $z=1$ halos with halo masses $10^{10.5} - 10^{12.5}\,\mathrm M_{\odot}$. 
The panels show snapshots at simulation time $t=6\,\mathrm{Gyr}$, after the inflow reached a quasi-steady state in all simulations (see below). The top panels show projected temperature and velocity vectors, while the bottom-panels show hydrogen density. The images are oriented edge-on with respect to the axis with the largest mass moment within $\Rvir$, and projections are made on a slice with thickness $0.1\,\Rvir$, using a volume average for $\nH$ and a mass-weighted geometric average for $T$. The solid and dashed circles mark $\Rvir$ and $R_\mathrm{sonic}$, respectively, where
$\Rsonic$ is the radius at which $\tcools=\tff$ (eqn.~\ref{eq:r_sonic}). We calculate $\tcools$ using the average density in each radial shell and assuming a temperature \changed{$T= (4/3)\Tvir$} (see eqn.~\ref{eq:t_cool}). 
The figure shows that
in the $10^{12.5}\msun$ simulation $\Rsonic$ is smaller than $0.1\Rvir$ so the gas is hot down to the galaxy scale. With decreasing halo mass  
$\Rsonic$ increases relative to $\Rvir$ as expected based on the analytic calculation above, implying that the galaxy is surrounded by cool turbulent gas. 
Note that the $10^{10.5}-10^{11}\msun$ halos have $\Rsonic\lesssim\Rvir$ in the simulations, in contrast with $\Rsonic\gtrsim\Rvir$ in the analytic calculation shown in Fig.~\ref{fig:CGM_phases}. This difference is the result of a factor of $2-3$ difference in the mean gas density near $\Rvir$ between the analytic estimate and the simulation. 

Fig.~\ref{fig:sim_img} demonstrates the critical role of $\Rsonic$ in turbulent CGM inflows. Within $\Rsonic$ a large fraction of the volume is filled with cool gas and the velocity field is highly turbulent. Large density fluctuations are also apparent, as expected when the turbulence is supersonic. 
In contrast, outside $R_\mathrm{sonic}$ the CGM is predominantly hot with a weaker turbulent velocity field and a more uniform density distribution. We explore these properties quantitatively in the following subsections.

\subsection{Enhancement of turbulence in CGM inflows}

In section~\ref{sec:turb} we argued based on \cite{Robertson12} that accretion flows drive turbulence due to the transfer of energy from the mean flow to the turbulent flow. In this subsection we explore how this process manifests in our CGM simulations. 

The top panel of Figure~\ref{fig:sigma_t_time} plots the turbulent velocity profile versus simulation time $t$ in the $10^{11.5}\,\mathrm M_{\odot}$, $z=1$ simulation (`m115z1', the simulation in the center panels of Fig.~\ref{fig:sim_img}).
The value of $\sigmat$ is calculated using eqn.~(\ref{eq:sigma_t}) on spherical radial shells spanning CGM radii of $0.05-2\Rvir$ with a relative width of $\delta r/r=0.07$. The panel shows ten different simulation times, from the initial conditions ($t=0$, yellow) to $t=7\,\mathrm{Gyr}$ (purple), while the horizontal gray line marks $V_\mathrm{c}$. Note that $\sigmat$ in the initial condition equals $\sigma_{\rm t,0}=23\kms$ only at large radii, due to the large scale of turbulence in the initial conditions (see Table~\ref{tab:sims_table}). 
The panel shows that within the first two Gyr, the turbulent velocity increases from the initial value of $\approx10\kms$ to $\sigmat\approx100\kms\approx\vc$ at radii within the sonic radius of $\approx30\,{\rm  kpc}$.  At larger radii beyond the sonic radius $\sigmat$ reaches a maximum value which decreases with radius. \changed{We emphasize that our simulations do not have explicit turbulence driving, so the only source of turbulent energy is the gravitational energy released in the inflow.} After $\sim3\,{\rm Gyr}$ the radial profile of $\sigmat$ remains constant with time, suggesting that the turbulent inflow is quasi-steady.

The \changed{second and third panels} of Fig.~\ref{fig:sigma_t_time} show the corresponding development of the mass-weighted average inflow velocity $\uin$ and the geometrically averaged sound speed $\csg$, where we use a geometric rather than arithmetic mean to give a better sense of the relative abundance of the cool and hot CGM
phases. The snapshots and radial shells are the same as in the top panel. 
At early times the inflow velocity rapidly increases from zero to values larger than $\vc$, with an inward accelerating $\uin$ profile indicating free-fall. This free-fall is a result of the rapid drop in $\csg$ and loss of pressure support, which is a consequence of rapid cooling at $r\lesssim\Rsonic$ where $\tcools<\tff$ \changed{(see eqns.~\ref{eq:t_cool}--\ref{eq:r_sonic})}. The inflow then starts to slow down, with $\uin$ first decreasing at inner radii and subsequently at larger radii, converging to $\uin\approx 50\kms\approx\vc/2$ at $r<\Rsonic$ and to lower values of $\uin$ at larger radii. This slowing down is coincident with the development of turbulence seen in the top panel, indicating that turbulence provides partial support against gravity within $\Rsonic$ and the flow transitions from free-fall to a turbulence dissipation-limited inflow. At the same time the sound speed converges to $\csg\ll\vc$  within $\Rsonic$ and to $\csg\approx\vc$ outside $\Rsonic$. After the transition to a dissipation limited inflow, the profiles of $\uin$ and $\csg$ remain constant,  demonstrating that the inflow becomes quasi-steady.  

\changed{The bottom panel of Fig.~\ref{fig:sigma_t_time} shows the evolution of the mean density profile. Mean densities initially decrease by a factor of a few, and then remain rather constant throughout the simulation. When the flow becomes quasi-steady, the mean density slope is $\langle\nH\rangle\sim r^{-2}$ at radii within the sonic radius  of $10-30\,{\rm kpc}$ and a somewhat shallower slope of $\sim r^{-1.5}$ at radii $50-200\,{\rm kpc}$ beyond the sonic radius.}

The quasi-steady inflow that develops in the simulation closely resembles the analytic solutions derived in section~\ref{sec:analytic}. \changed{At radii  of $10-30\,{\rm kpc}$ within the sonic radius, Fig.~\ref{fig:sigma_t_time} indicates that $\sigmat\approx\vc$ and $\uin\sim\sigmat$, with both velocities remaining constant with radius and $\langle n_{\rm H}\rangle\propto r^{-2}$. This behavior is consistent with the turbulence-dominated inflow solution (eqn.~\ref{eq:cool_solution}).} At larger radii outside $\Rsonic$ the flow is dominated by thermal pressure, and again $\uin\sim\sigmat$ with both velocities increasing inward as expected in a hot turbulent inflow (eqn.~\ref{eq:cooling_flow}). In general, the $\ur$ and $\sigmat$ profiles are similar at all radii up to a constant of order unity, as expected due to the balance of turbulence dissipation and turbulence driving by the inflow (see Fig.~\ref{fig:partial_time_derivative} and section~\ref{sec:turb}). The ratio $\sigmat/\ur$ in the simulations is further explored in section~\ref{sec:eta} below.  

\changed{
Fig.~\ref{fig:sigma_t_time_iso} plots the development of the $\sigmat$ profile in the isothermal simulation `m115z1\_a', which has the same parameters and initial conditions as the simulation used in Fig.~\ref{fig:sigma_t_time} except for a gas temperature equal to the equilibrium temperature (eqn.~\ref{eq:cs_eq}). 
As in the radiative simulation, at inner CGM radii the turbulent velocity rapidly increases from the initial value of $\sigmat\approx10\kms$, reaching $\sigmat\approx100\kms\approx\vc$ within a couple of Gyr. At outer radii of $r>60\,{\rm kpc}$ the value of $\sigmat$ develops an inward increasing profile. 
The inflow that develops thus has two parts, a turbulence-dominated inflow at inner radii similar to the radiative solution, and a cool free-fall at outer radii, in contrast with a hot inflow at outer radii in the radiative solution. 
The similar development of turbulent velocity in the isothermal and radiative simulations indicates that turbulence is driven by `adiabatic heating', rather than by thermal effects. }

Figure~\ref{fig:velocity_radial_profiles} shows how the quasi-steady turbulent inflow solution depends on halo mass, using three radiative simulations of halos at redshift $z=1$, with halo masses of $M_{\text{halo}}=10^{10.5}\,\mathrm M_{\odot}$ (left),  $10^{11.5}\,\mathrm M_{\odot}$ (middle) and $10^{12.5}\,\mathrm M_{\odot}$ (right), corresponding to the first, third and fifth columns in Fig.~\ref{fig:sim_img}. The panels show radial profiles of mass-weighted inflow velocity (blue), turbulent velocity (black), and mass-weighted geometrically averaged sound-speed (red). The averaging is performed both over solid angle and over time, on snapshots with times $5-6.5\,{\rm Gyr}$ which correspond to $4-5$ times the free-fall time measured at $\Rvir$. At these times  the turbulent inflow is quasi-steady at radii smaller than the cooling radius (noted on top, calculated using eqn.~\ref{eq:Rcool}), so the time-averaging smooths over the variability of the turbulent flow. 

In the $10^{12.5}\msun$ simulation (right panel of Fig.~\ref{fig:velocity_radial_profiles}) the value of $\Rsonic$ is $<0.05\Rvir$, at the galaxy scale. This is in contrast with the $10^{11.5}\msun$ simulation shown in the middle panel and in Fig.~\ref{fig:sigma_t_time}, where the sonic radius is on CGM scales with $\Rsonic=0.4\,\Rvir$. 
The properties of the turbulent CGM inflow in both simulations are however similar relative to the sonic radius. 
At $r>\Rsonic$ we find $\csg\approx\vc$ so the gas is hot, while $\sigmat$ and $\uin$ are smaller than $\vc$ and increase inward. This is the slow cooling limit discussed above: a thermal pressure-dominated inflow with mild turbulence (section~\ref{sec:hot}).
At $r<\Rsonic$ we find flat $\sigmat$ and $\uin$ profiles with $\ur\sim\sigmat\approx\vc$, while the mean sound speed is significantly lower ($\csg\ll\vc$). This is the rapid cooling limit where turbulence dominates (section~\ref{sec:cool}).

The $10^{10.5}\msun$ simulation shown in the left panel of Fig.~\ref{fig:velocity_radial_profiles} has $\Rsonic/\Rvir\approx0.7$, larger than the other two simulations. Within this radius we again find $\ur\sim\sigmat\approx\vc$ and $\csg\ll\vc$. At radii larger than $\Rsonic$ the panel shows that $\uin$ is high and close to $\vc$, while $\sigmat$ increases inwards as $\sim r^{-1}$ and $\cs\lesssim\vc$. The simulation is therefore in the turbulence-dominated regime at $r<\Rsonic$ and in the free-fall regime (eqn.~\ref{eq:cold_flow}) at $r>\Rsonic$. The turbulence-dominated regime appears at inner CGM radii in all our $\leq10^{12}\msun$ simulations where $\Rsonic$ is larger than the galaxy scale.

The $10^{11.5}\msun$ simulation shown in the middle panel of Fig.~\ref{fig:velocity_radial_profiles} exhibits a drop in $\uin$ at radii $r\lesssim7\,{\rm kpc}$, indicating the inflow halts at small radii. This is a result of the mean specific angular momentum $\langle j\rangle$ reaching $\approx V_\mathrm{c} r$ at these radii, so rotation provides support against gravity. This angular momentum is a stochastic result of the initial turbulent velocity field, since we did not explicitly choose a preferred rotation axis in the initial conditions. The value of $\langle j\rangle$ is small relative to $\vc r$ in all other radii and simulations shown in Fig.~\ref{fig:velocity_radial_profiles}, indicating that rotational support is negligible.

\begin{figure}
	
 \centering
	\includegraphics[width=\columnwidth]{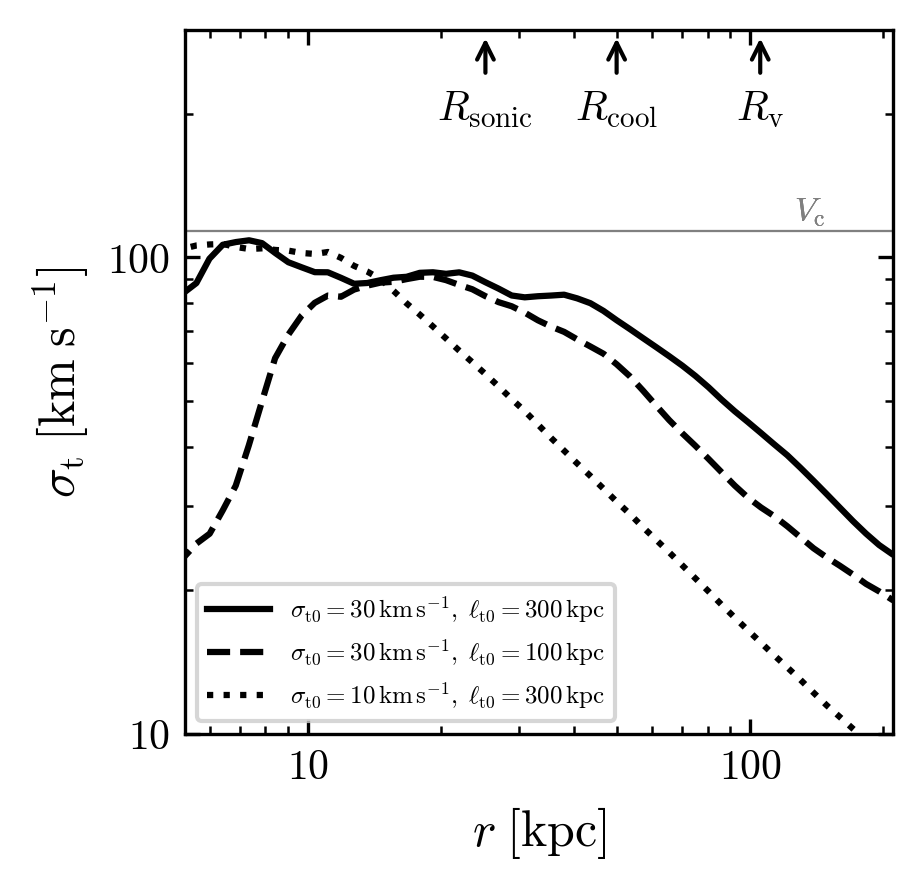}
    \vspace{-0.5cm}
    \caption{Turbulent velocity profiles of CGM accretion flows with different initial conditions, in a $10^{11.5}\msun$ halo at $z=1$. 
    Solid line plots the simulation with fiducial initial conditions. Dashed and dotted lines plot simulations with lower initial turbulent scale and lower initial turbulent velocity, respectively.  
    In all cases $\sigmat$ increases inwards until reaching $\sim \vc$ in the inner CGM. 
    }
    \label{fig:different_ics_vel}
\end{figure}

\begin{figure*}
	
    \centering
    \includegraphics[width=\textwidth]{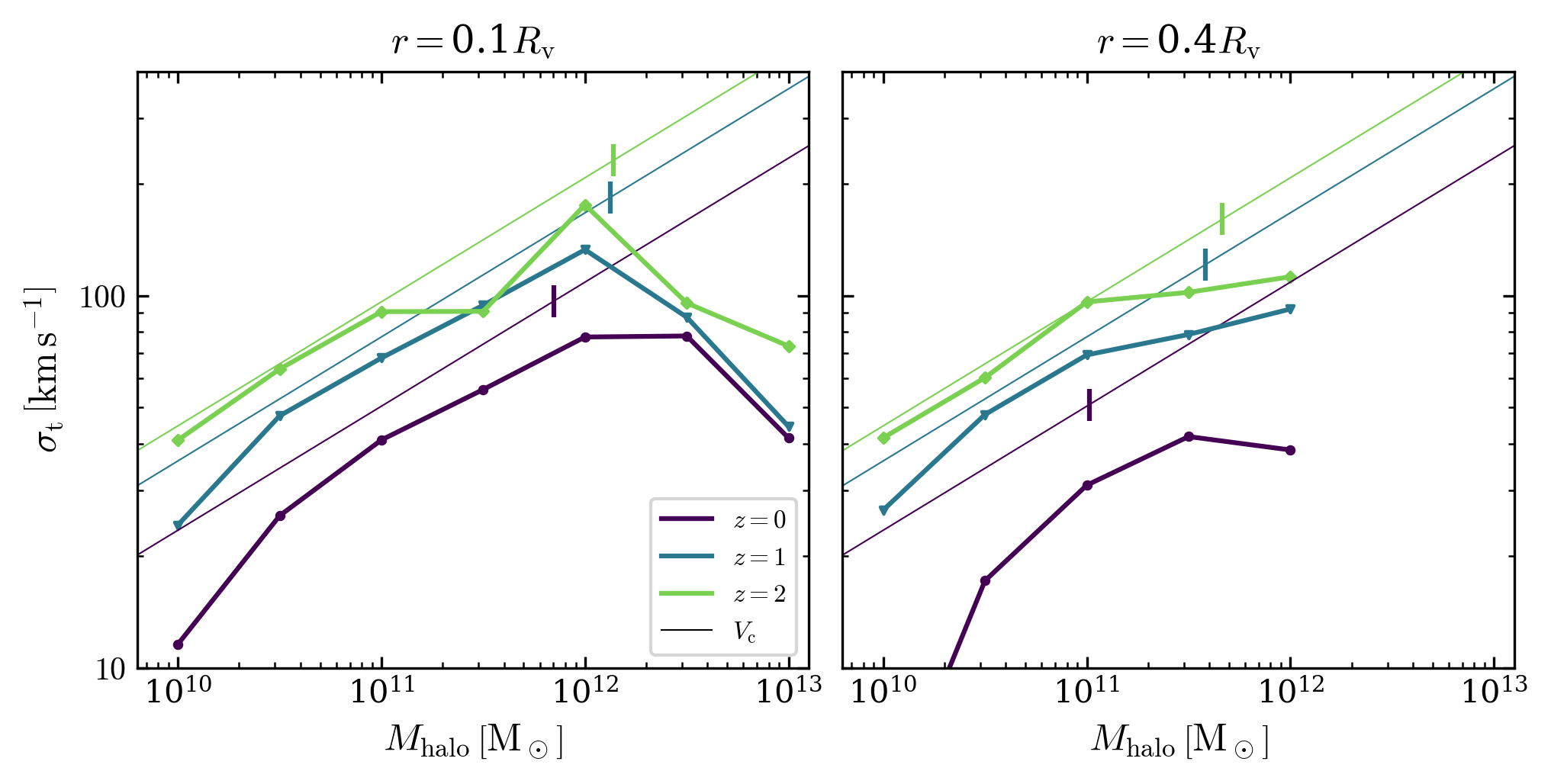}
    \vspace{-0.5cm}
    \caption{Predicted turbulent velocity in CGM accretion flows versus halo mass, 
    for different redshifts (different colors) and for different radii (different panels, noted on top). Thin lines show $V_\text{c}$. The `|' symbols mark the mass at which $R_{\rm sonic}$ equals the radius shown in the panel. We find $\sigma_t\approx V_\text{c}$ at low masses  where cooling is rapid relative to free-fall, while $\sigma<\vc$ at higher masses where cooling is slow.  
    }
    \label{fig:sigma_t_M_vir}
\end{figure*}

Figure~\ref{fig:different_ics_vel} explores how our conclusion that $\sigmat\approx\vc$ at inner CGM radii depends on the parameters of the initial turbulent velocity field. 
The figure shows $\sigmat$ profiles in simulations with different initial turbulent velocity $\sigma_{{\rm t},0}$ and initial turbulent velocity scale $l_{{\rm t},0}$ (defined in section~\ref{sec:ICs}), for a $10^{11.5}\msun$ halo at $z=1$ (simulations m115z1, m115z1\_b and m115z1\_c in Table~\ref{tab:sims_table}). We plot the profiles at the same simulation times as shown in Fig.~\ref{fig:velocity_radial_profiles}. 
In all cases the turbulent velocity profiles increase inward and reach $\sigmat\approx\vc$ at small radii. 
In the two simulations with $\sigma_{{\rm t},0}=30\kms$ the turbulent velocity increases inwards as $\sim r^{-0.5}$ as expected in thermal pressure-dominated gas with $\tdiss\approx \tflow$ (eqn.~\ref{eq:cooling_flow_turb}), reaching and remaining $\approx\vc$ at radii within $\Rsonic$.  In the simulation initialized with a lower turbulent velocity of $\sigma_{{\rm t},0}=10\kms$, the turbulence velocity profile has a steeper slope closer to $r^{-1}$ as expected when $\tdiss\gg\tflow$  (eqn.~\ref{eq:sigmat free-fall}), so it saturates at $\vc$ only within $\Rsonic$. 
The simulation initialized with a lower turbulent scale of $\ell_{{\rm t},0}=100\, \mathrm{kpc}$ (dashed line) exhibits a drop in $\sigmat$ at radii $r<10\,{\rm kpc}$. This is a result of the drop in $\uin$ due to rotation that provides support against gravity as explained in the previous section. Except for this difference this simulation is similar to the simulation with the fiducial turbulent scale.

\subsection{Predicted turbulence velocity versus halo mass, redshift and radius} 

The predicted turbulent velocities in CGM inflows are plotted in Figure~\ref{fig:sigma_t_M_vir} as a function of halo mass, redshift, and radius, based on all simulations listed in Table~\ref{tab:sims_table}. Panels show different radii relative to $\Rvir$, while line color corresponds to different redshifts as noted in the legend. The `|' symbols marks the mass at which $\Rsonic$ equals the radius shown in the panel, which is $\approx10^{12}\msun$ in most cases. At halo masses below this threshold the flow is in the cool supersonic regime at the shown radius, while at halo masses above the threshold the flow is in the hot subsonic regime at this radius.  We find $\sigma\approx\vc$ in halo masses below this threshold and above $10^{10}\msun$, as expected from equation~(\ref{eq:cool_solution}) and as previously seen in Figs.~\ref{fig:sigma_t_time} -- \ref{fig:velocity_radial_profiles}, indicating that cool supersonic inflows are dominated by turbulence. The somewhat lower $\sigmat\approx0.5\vc$ in the ($10^{11.5}\msun, z=2$) simulation at $r=0.1\Rvir$ is due to rotation support in the shown snapshot. The ubiquity of $\sigmat\approx\vc$ in the inner CGM of $10^{10}-10^{12}\msun$ halos shown in Fig.~\ref{fig:sigma_t_M_vir} is a main result of this work. 

In the ($10^{10}\msun, z=0$) simulation shown in Fig.~\ref{fig:sigma_t_M_vir} the value of $\sigmat$ is smaller than $\vc$, as also is the case to a lesser extent in the ($10^{10}\msun, z=1$) simulation.  This is likely since $\vc$ is close to $c_{\rm s, eq}$ in these halos, so the pressure support is substantial even for gas at the equilibrium temperature, and hence the rapid cooling limit does not apply (see section~\ref{sec:inflows}). At large halo masses above the `|', Fig.~\ref{fig:sigma_t_M_vir} shows that the turbulent velocity stops tracing $\vc$.  This follows since at these halo masses and radii the CGM is in the hot, slow-cooling regime where $\sigmat \sim r/\tcool$ (eqn.~\ref{eq:cooling_flow_turb}).

\newcommand{\machtavg}{\langle\mathcal{M}_{\rm t}\rangle}

Figure~\ref{fig:velocities_cool2ff} plots $\sigmat/\vc$ versus $\tcools/\tff$, the dimensionless ratio that is expected to set $\sigmat/\vc$ (see section~\ref{sec: turb inflows}).  
Each marker corresponds to a different radial shell in the radial range $0.05\,\Rvir<r<\Rvir$, averaged over snapshots with $4<t/\tff(\Rvir)<5$ at which quasi-steady state has been achieved in our simulations. We use only simulations with $V_\mathrm{c}>25\, \mathrm{km\, s^{-1}}$ so virial temperature gas is far from the equilibrium temperature. Markers are colored according to the mean Mach number in the shell, calculated as 
\begin{equation}\label{eq:mach_t sim}
    \machtavg = \frac{\sigmat}{\csg} ~.
\end{equation} 
In radial shells with $\tcools<\tff$ we find $\sigmat\approx\vc$ and supersonic Mach numbers, i.e., the flow is in the cool, turbulent-dominated regime. 
When $\tcools>\tff$ the ratio $\sigmat/\vc$ decreases with increasing $\tcools/\tff$ and the turbulence is subsonic, as expected in the hot inflow regime (eqn.~\ref{eq:cooling_flow_turb}). 
The plot exhibits a narrow distribution of $\sigmat/\vc$ at a given $\tcools/\tff$, indicating that this ratio is indeed the main parameter which sets $\sigmat/\vc$ in our simulations. The dependence of $\sigmat$ on halo mass, redshift and radius shown in Fig.~\ref{fig:sigma_t_M_vir} is mainly a result of how these parameters affect $\tcools/\tff$.

\subsection{Dissipation coefficient $\eta$}\label{sec:eta}

The turbulence dissipation coefficient $\eta$ (eqn.~\ref{eq:dissipation})  sets the ratio between turbulent and inflow velocity (eqn.~\ref{eq:sigmat_ur}).  
Thus, to estimate $\eta$ in our simulations we plot in Figure~\ref{fig:sigma_t_over_u_r} the ratio $\sigmat/\uin$, colored by the turbulent Mach number defined in eqn.~(\ref{eq:mach_t sim}). Each marker corresponds to a different radial shell in the simulations, at times and radii where the flow is quasi-steady as in Fig.~\ref{fig:velocities_cool2ff}. 
The figure shows that $\sigmat/\uin$ approaches a constant at inner CGM radii, with the asymptotic value decreasing with increasing Mach number.  In the supersonic regime we find 
$\sigmat/\uin\approx1.5$ at inner CGM radii, which corresponds to $\eta_\text{supersonic}\approx 0.7$ using eqn.~(\ref{eq:sigmat_ur}) and $p\approx0$. In the subsonic regime, Fig.~\ref{fig:sigma_t_over_u_r} shows $\sigmat/\uin\approx3$ which corresponds to $\eta_\text{subsonic}\approx 0.2$ since $p\approx-0.5$. 

Similar values of $\eta_\text{subsonic}\approx0.2-0.25$ and $\eta_\text{supersonic}\approx1$ were found by \citet{Hennebelle21} in a different simulation setup of self-gravitating collapsing clouds. Our simulations thus support their conclusion that the dissipation of turbulent energy is faster, relative to the eddy turnover time, in supersonic gas. \changed{This agreement is despite that turbulence is driven explicitly in their simulations, while in the simulations herein turbulence is driven by accretion. }

The small ratios of $\sigmat / \uin\ll1$ seen in Fig.~\ref{fig:sigma_t_over_u_r} at outer CGM radii originate in simulations of low mass halos ($\leq10^{11}\msun$). The low ratios indicate that adiabatic heating is not balanced by dissipation, consistent with the result that the inflow is in free-fall at these halo masses and radii (Fig.~\ref{fig:velocity_radial_profiles}, left panel). The dominance of adiabatic heating also explains why $\sigmat/\uin$ increases rapidly inward when $\sigmat\ll\uin$ in  Fig.~\ref{fig:sigma_t_over_u_r}, since heating dominance implies that $\sigmat\propto r^{-1}$  while $\uin$ increases inward only logarithmically (eqn.~\ref{eq:sigmat free-fall}).

\subsection{Turbulence anisotropy}\label{sec:anisotropy}

\begin{figure}
    \centering
    \includegraphics[width=\columnwidth]{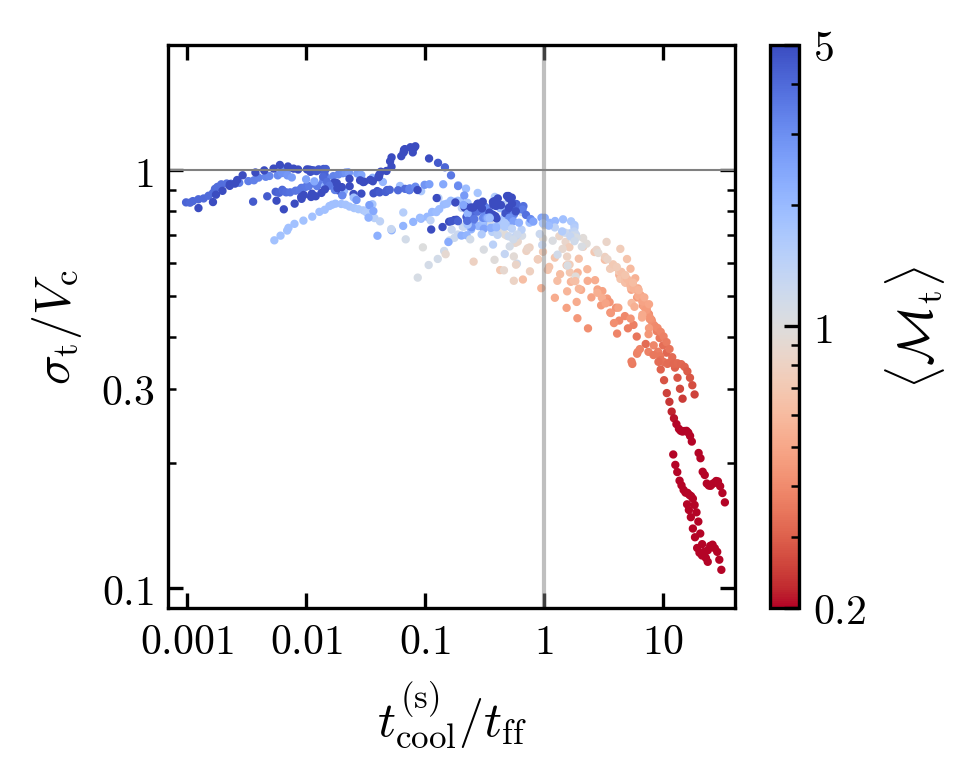}
    \vspace{-0.5cm}
    \caption{Turbulence velocity versus cooling to free-fall time ratio. \changed{The cooling time is calculated assuming the gas is hot (eqn.~\ref{eq:t_cool}).}
    Each marker corresponds to a different radial shell at $0.05\Rvir<r<\Rvir$ in snapshots where the flow is quasi-steady, from all simulations listed in table~\ref{tab:sims_table} with $V_\mathrm{c}>25\kms$. Color denotes the mean turbulent Mach number in the shell (eqn.~\ref{eq:mach_t sim}).  Turbulence is supersonic when $\tcools<\tff$ and subsonic when $\tcools>\tff$. A narrow distribution of $\sigmat/\vc$ is apparent at a given $\tcools/\tff$, indicating that this ratio is the main parameter setting $\sigmat/\vc$ in CGM inflows. 
    }
    \label{fig:velocities_cool2ff}
\end{figure}

\begin{figure}	
	\includegraphics[width=\columnwidth]{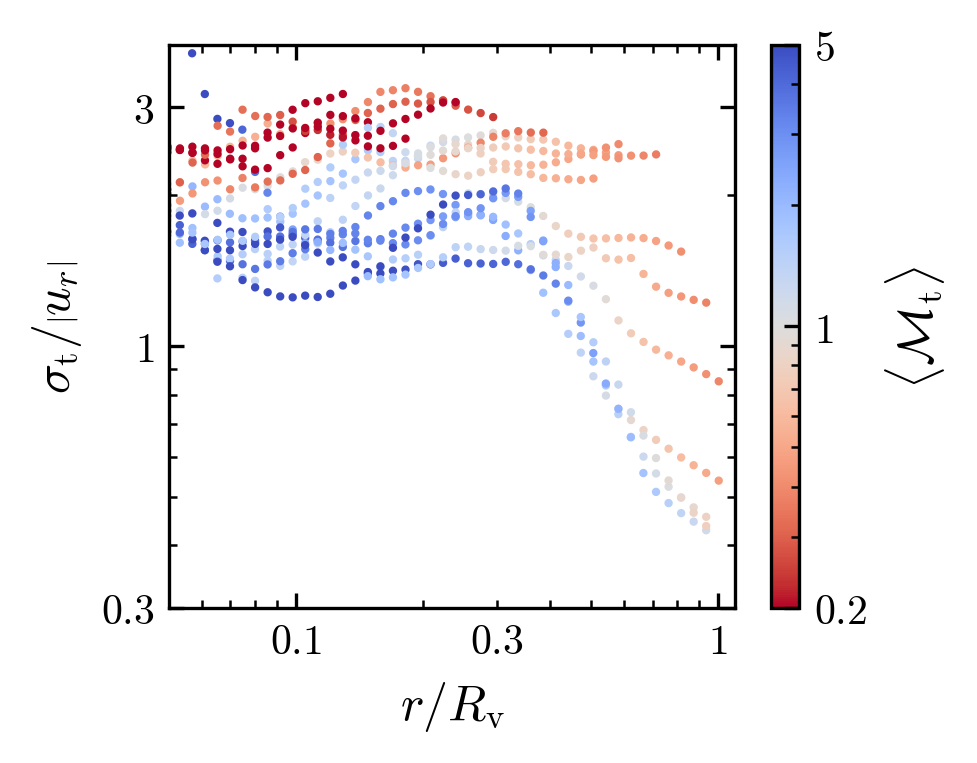}
        \vspace{-0.5cm}
    \caption{Ratio of turbulent velocity to inflow velocity versus radius, in the simulations listed in Table~\ref{tab:sims_table} at times when the flow is in quasi-steady state. Color denotes turbulent mach number.  
    The ratio tends to be independent of radius at inner CGM radii, spanning between $\sigmat/\uin\approx 1.5$ for supersonic gas and $\sigmat/\uin \approx 3$ for subsonic gas.}
    \label{fig:sigma_t_over_u_r}
\end{figure}

\begin{figure}	
	\includegraphics[width=\columnwidth]{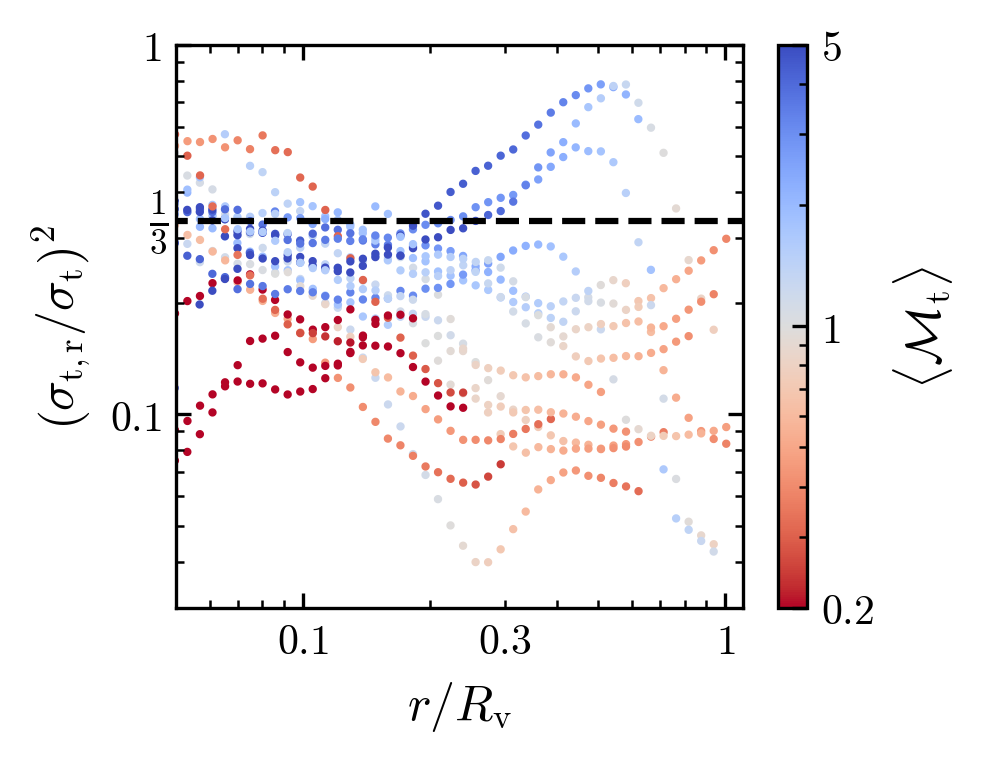}
    \vspace{-0.5cm}
    \caption{Fraction of turbulent energy in the radial component, in the simulations listed in Table~\ref{tab:sims_table} at times when the flow is steady. Color denotes turbulent Mach number. 
The fraction is $\approx1/3$ when turbulence is supersonic at $\lesssim0.3\Rvir$, indicating roughly isotropic turbulence.  When turbulence is subsonic the fraction is typically $<1/3$, indicating suppressed radial modes. 
}
    \label{fig:sigma_rr_to_sigma_t}
\end{figure}

The radial gravitational field of the halo induces a preferred direction in our simulations, which is typically absent from turbulent box simulations mimicking a patch of the ISM.  Figure~\ref{fig:sigma_rr_to_sigma_t} explores whether this effect induces anisotropy in CGM turbulence driven by inflows, by plotting the square of the ratio between the radial component $\sigma_{{\rm t},r}$ and the total turbulent velocity $\sigmat$. The isotropic value of $1/3$ is marked with a horizontal line. As in Figs.~\ref{fig:velocities_cool2ff} -- \ref{fig:sigma_t_over_u_r} each marker corresponds to a different radial shell and snapshot in the simulations at times and radii where the flow is quasi-steady, with marker color corresponding to the Mach number in the shell. The figure shows that turbulence is roughly isotropic ($\sigma_{{\rm t},r}^2/\sigmat^2\approx0.25-0.3$) at small radii with supersonic turbulence. In contrast turbulence is dominated by tangential modes, with $\sigma_{{\rm t},r}^2/\sigmat^2$ equal to  $0.15$ or lower when turbulence is subsonic.  

A potential origin for the anisotropy in the subsonic regime is buoyancy damping of the radial component, which is expected when the Froude number is small \citep[e.g.][]{Mohapatra23}. The Froude number is defined as ${\rm Fr} = \sigmat/N$ where $N^2=\vc^2(\alpha_P/\gamma-\alpha_\rho)$ is the Brunt-V\"ais\"al\"a frequency and $\alpha_P$ and $\alpha_\rho$ are the power-law indices of the pressure and density profiles. Using $\alpha_P=\alpha_\rho=-1.5$ (eqn.~\ref{eq:cooling_flow T and rho}) we get $N=0.8\vc$, so the Froude number is indeed small in the subsonic regime where $\sigmat\ll\vc$.

\subsection{Density distribution}\label{sec:density sim}

\begin{figure*}
	
 \centering
	\includegraphics[width=\textwidth]{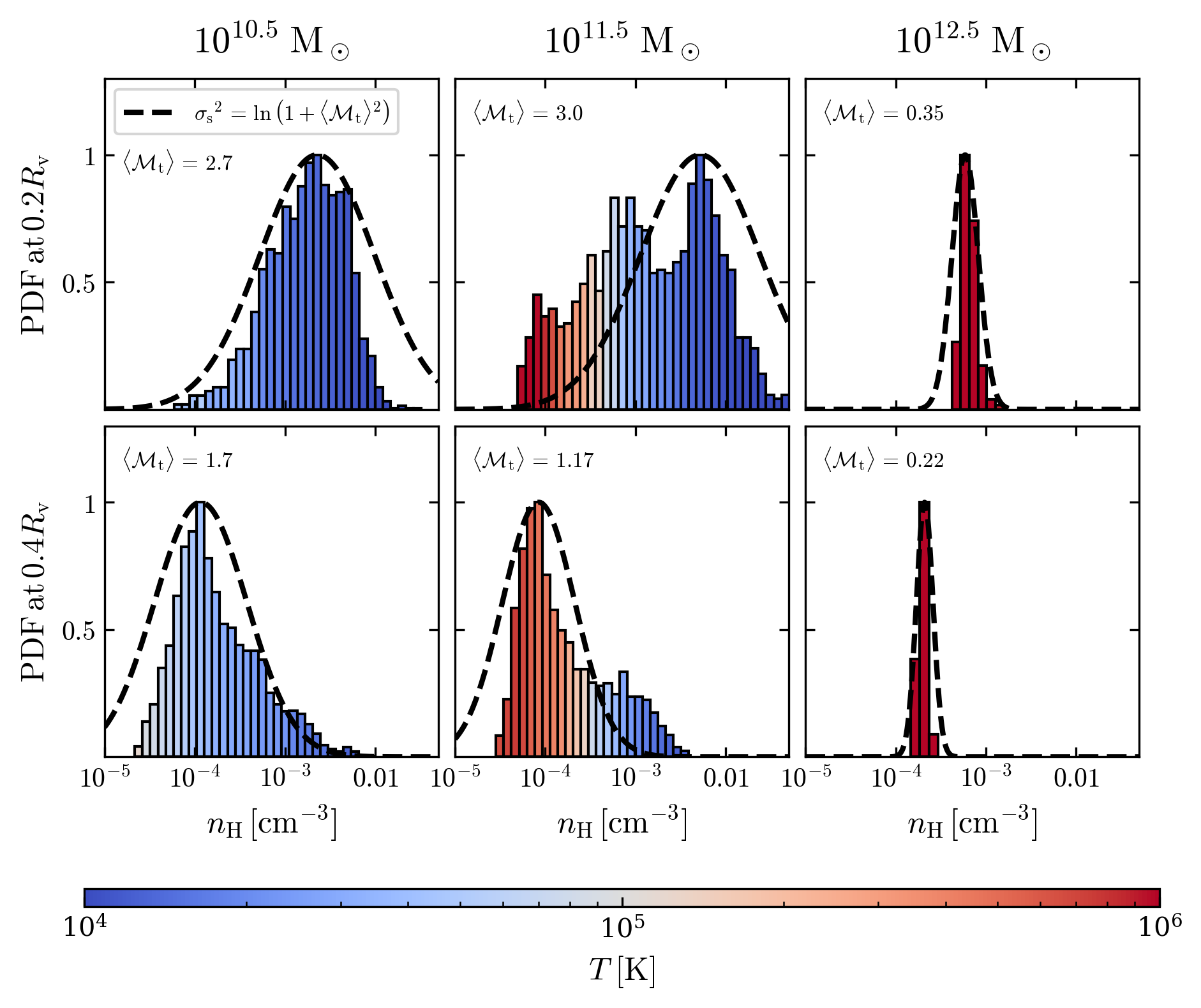}
    \vspace{-0.5cm}
    \caption{Mass weighted gas density distributions in the three simulations from Fig.~\ref{fig:velocity_radial_profiles}, in two radial shells (top:\ $0.2R_\mathrm{v}$, bottom:\ $0.4R_\mathrm{v}$). Color marks the geometric average temperature in each density bin.
    Note the prominence of the cool phase in the top-middle, top-left, and bottom-left panels, which correspond to radii within the sonic radius.    
Dashed lines show the expected lognormal distributions (eqn.~\ref{eq:sigma_log_rho}) for the mean Mach number calculated in the shell (eqn.~\ref{eq:mach_t sim}, noted in panels) \changed{and assuming $b=1$. The density dispersion of the dominant phase increases with mean Mach number as expected, with values suggesting $b$ somewhat lower than unity.} 
}
    \label{fig:density_PDF}
\end{figure*}

In homogeneous isothermal turbulence the density dispersion is set by the turbulent Mach number  as discussed in section~\ref{sec:density_distribution}. Here we test to what extent we can apply this result to a stratified and radiative CGM. This predicted density distribution can be tested against UV absorption observations which provide measurements of gas columns of different ions, since the gas density of the cool CGM phase determines its ionization state \citep[e.g.,][]{Tumlinson17, Faerman25, Kakoly25}.

Figure~\ref{fig:density_PDF} shows density distributions of the three halos presented in Fig.~\ref{fig:velocity_radial_profiles}  at two radii $r=0.2\,\Rvir$ (top) and $r=0.4\,\Rvir$ (bottom). Each histogram includes all resolution elements in a thin shell with a width of $\delta r/r=0.1$, which avoids the effect of radial gradients on density dispersion. Color indicates the geometric mean temperature of gas in a given density bin. As suggested by Fig.~\ref{fig:sim_img}, the cool gas mass fraction increases with decreasing halo mass and radius, due to the decrease in $\tcools/\tff$. In the low- and high-mass simulations a single phase dominates and the distribution is roughly lognormal. The lognormal is narrow when the gas is hot and turbulence is subsonic ($\sigma_s=0.2$ at $0.2\Rvir$ and $\sigma_s=0.1$ at $0.4\Rvir$), while the lognormal is wide when the gas is cool and supersonic ($\sigma_s=1.0$ at both radii). A bi-modal distribution is apparent in the middle panels where the mass is comparable between the two phases. 

We compare each density histogram in Fig.~\ref{fig:density_PDF} to a lognormal with a mode equal to the peak of the distribution and a width equal to the expected $\sigma_s$ based on eqn.~(\ref{eq:sigma_log_rho}), assuming a turbulent Mach number calculated using eqn.~(\ref{eq:mach_t sim}) and $b=1$ as expected for compressive driving (dashed lines).  
\changed{
The expected lognormal width in the low mass halo are $\sigma_s=1.5$ at $r=0.2\Rvir$ and $\sigma_s=1.2$ at $r=0.4\Rvir$. In the high mass halo the corresponding values are $\sigma_s=0.3$ at $r=0.2\Rvir$ and $\sigma_s=0.2$ at $r=0.4\Rvir$. These expected values have a similar trend with Mach number as the dispersion in the simulation, suggesting that turbulence is indeed the main driver of the simulated density distribution. However, the expected $\sigma_s$ assuming $b=1$ are $20-50\%$ higher than the simulated values. The values of $b$ which best-fit the simulation results are in the range $0.5-0.8$.
}

\subsection{Velocity structure function (VSF)}\label{sec:VSF}
\begin{figure}

\includegraphics[width=\columnwidth]{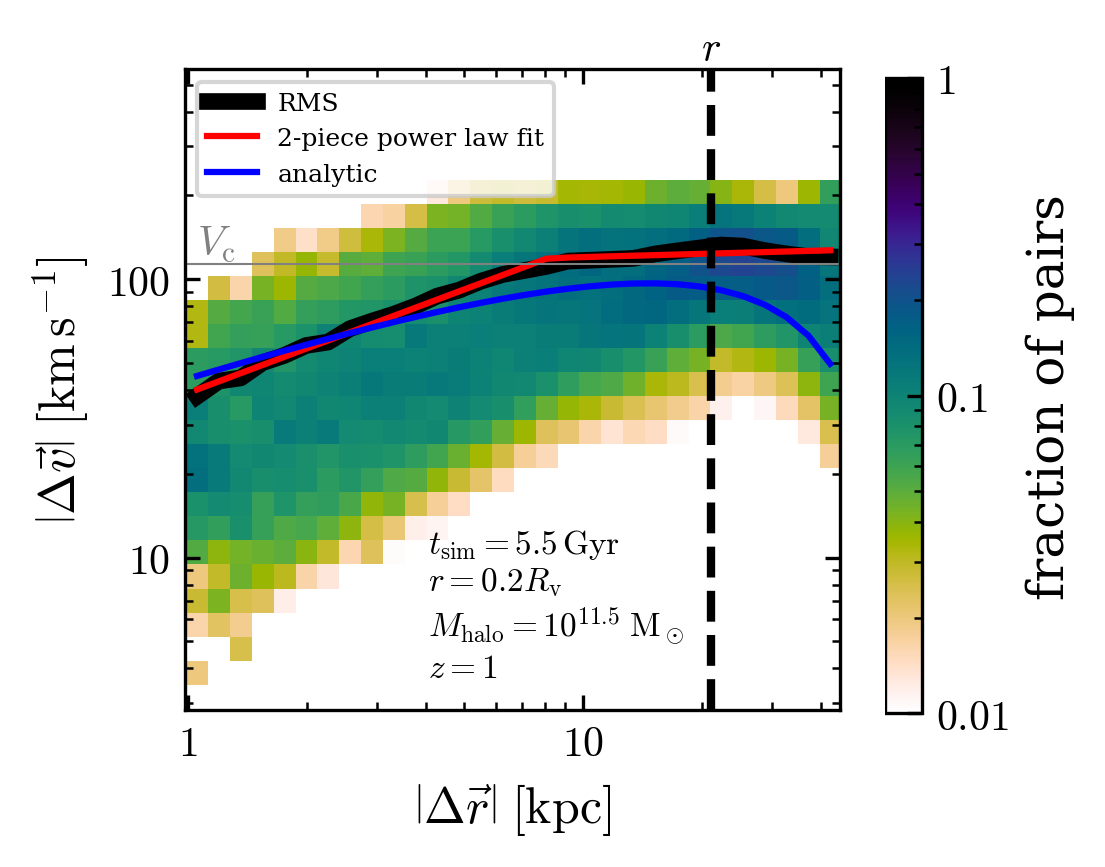}
\vspace{-0.5cm}
\caption{Velocity structure function (VSF) of a shell at $r=0.2\Rvir$ in the $10^{11.5}\msun,\ z=1$ simulation, after a quasi-steady state has been achieved. The background column-normalized histogram denotes the distribution of velocity differences between particles with a given physical separation $|\Delta \vec{r}|$. Black line traces the RMS velocity for each separation, red line denotes a two-piece power law fit to the RMS, while the blue plots the analytic VSF prediction (eqn.~\ref{eq: VSF with heating}). The shell radius and $\vc$ are also marked.
The velocity differences increase with separation, reaching $\approx\vc$ at a scale comparable to the shell radius.
}
    \label{fig:VSF_example}
\end{figure}

\newcommand{\sep}{|\Delta \vec{r}|}
\newcommand{\veldiff}{|\Delta \vec{v}|}

In this subsection we calculate VSFs in our simulations, and compare to those expected in standard turbulence spectra. 
VSFs at a specific radius and time in the simulations are calculated by randomly selecting $5000$ pairs of resolution elements within radial shells of width $\delta r/r=0.1$. For each pair `1' and `2' we calculate the Cartesian distance $\sep$ which we call the `separation', and the velocity difference $\veldiff$ in Cartesian coordinates after subtracting the mean radial and tangential velocities of the shell:
\begin{equation}
    \veldiff = \left|\vec{u}_1' -\vec{u}_2'\right| ~,
\end{equation}
where 
\begin{equation}
\vec{u}_i'=\vec{u}_i-\langle u_r\rangle\hat{r}- \langle\vec{u}_{||}\rangle \changed{~~~~\left(i\in\{1,2\}\right)}
\end{equation}
The mean tangential velocity $\langle\vec{u}_{||}\rangle$ is calculated as in \cite{Murray17}:
\begin{equation}
    \vec{u}_{||}=\left(I^{-1}\vec{L}\right)\times \vec{r} ~.
    \label{eq:tangent_vel}
\end{equation} 
where $I_{ij}=\langle\delta_{ij}r^2 - x_i x_j\rangle$ is the moment of inertia tensor of the shell, $x_i$ and $x_j$ are the Cartesian coordinates of a resolution element, and $\delta_{ij}$ is the Kronecker delta.

Figure~\ref{fig:VSF_example} plots column-normalized histograms of $\veldiff$ versus $\sep$ for an example shell 
at $0.2\Rvir$ in the $t=5.5\,{\rm Gyr}$ snapshot of the $M_\mathrm{halo}=10^{11.5}\msun$, $z=1$ simulation. 
The black line traces the RMS of the velocity difference versus  separation, which we refer to here as the VSF. The red line plots a two-piece broken power law fit to the VSF. 
The maximum of the VSF is $\approx\vc$ and the slope at lower separations is 0.55, comparable to the Burgers power law index of $0.5$ expected in supersonic turbulence.
The distribution of velocity differences at a given separation is wide, with a standard deviation of $30-100\kms$. 
\begin{figure*}
	
 \centering
\includegraphics[width=\textwidth]{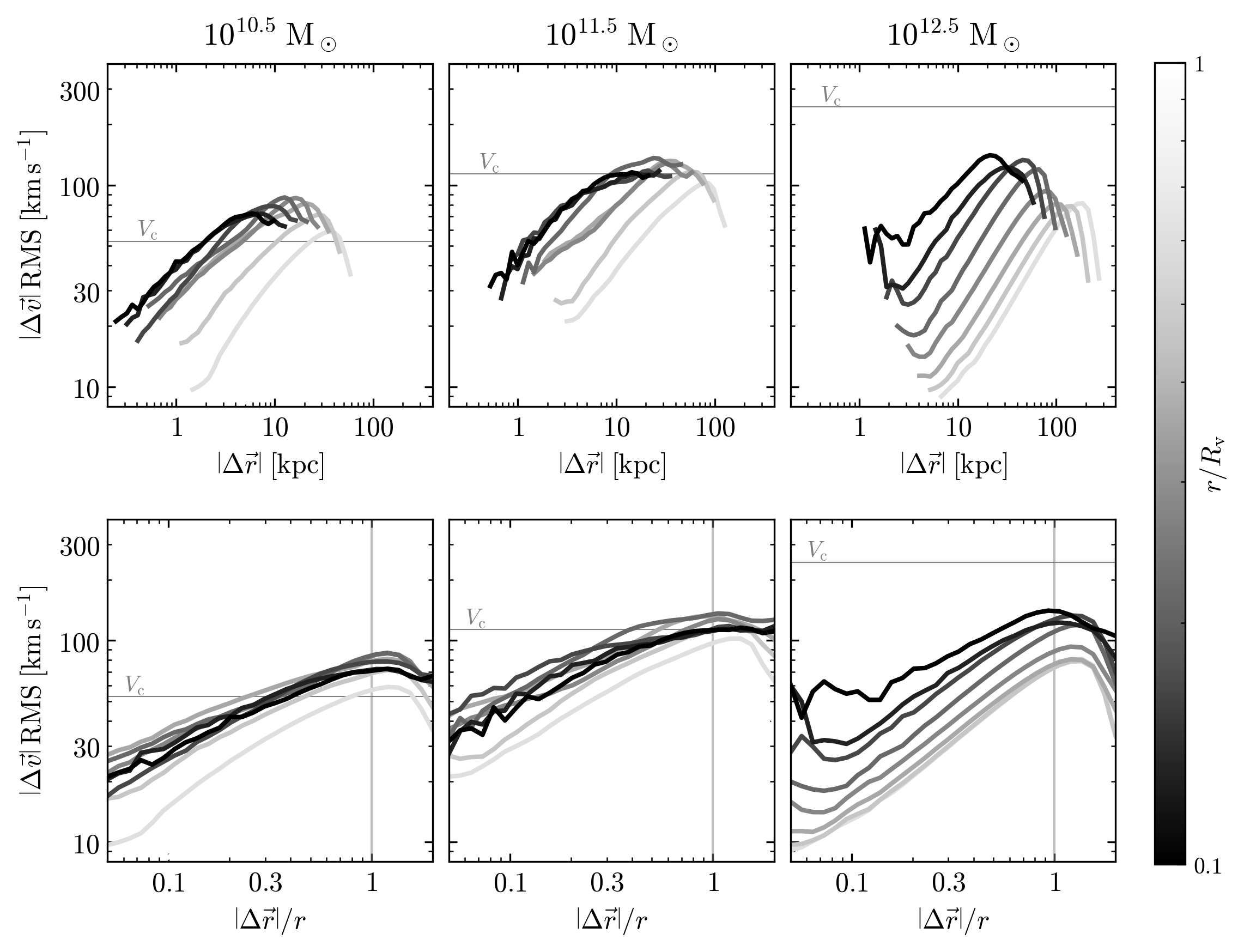}
\vspace{-0.5cm}
\caption{Top: Velocity structure functions at different radii in the simulations shown in Fig.~\ref{fig:velocity_radial_profiles}, after a quasi-steady state has been achieved. The greyscale indicates shell radius in units of the $\Rvir$.
The VSF normalization tends to increase with decreasing radius, and the change in slope occurs at smaller separations. 
Bottom: Similar to top panels,  with separation normalized by shell radius. The change in VSF slope occurs at a separation comparable to the shell radius. 
}
    \label{fig:VSF_radii}
\end{figure*}

\changed{The blue line plots the VSF predicted by eqn.~(\ref{eq: VSF with heating}), which at separations $l\sim r$ is similar to the VSF in the simulation, while at smaller separations is somewhat flatter than the simulation. } 

Figure~\ref{fig:VSF_radii} plots the VSFs in the three simulations shown in Fig.~\ref{fig:velocity_radial_profiles}, using the $t=5.5\,{\rm Gyr}$ snapshots when the inflows are quasi-steady. Top panels show the VSFs with the separation in physical units, and bottom panels show separation normalized by the shell radius. The grayscale indicates the radius of the shell relative to $\Rvir$. Calculations of the VSFs were done using $5000$  resolution elements in the shell, or the total number of resolution elements in the shell if there is less than $5000$. The figure demonstrates that the maximum of the VSFs increases with decreasing radius, reaching $\vc$ or more in the $10^{10.5}$ and $10^{11.5}\msun$ simulations. Since the maximum of the VSF is approximately equal to $\sigmat$, this result is consistent with the $\sigmat$ profile analysis in Figs.~\ref{fig:sigma_t_time} -- \ref{fig:velocity_radial_profiles}.

Fig.~\ref{fig:VSF_radii} shows also that for smaller radii the change in VSF slope occurs at smaller separations. This trend is further explored in the bottom panels, which plots the VSF as a function of the separation normalized by the radius of the shell. The change in slope in the VSF occurs at a separation that is comparable to the radius,  indicating that the driving scale in inflow-driven turbulence is indeed the shell radius, as argued in section~\ref{sec:VSF analytic}.

\begin{figure*}
	
 \centering
	\includegraphics[width=\textwidth]{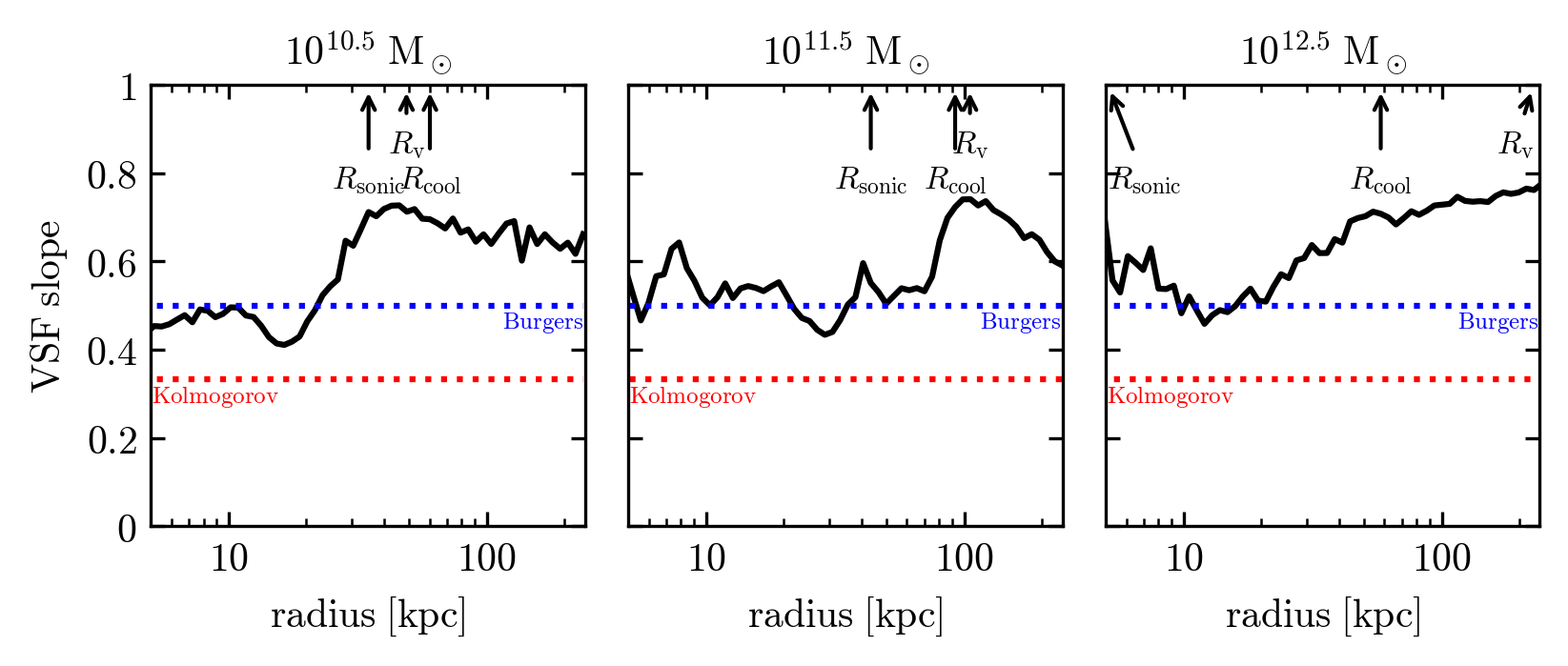}
    \vspace{-0.5cm}
    \caption{VSF slope versus radius for the three simulations shown in Fig.~\ref{fig:VSF_radii}. The slope is calculated from the two-piece power-law fit to the VSFs, using the slope at small separations. Blue and red lines mark the Kolmogorov and Burgers indices expected for unstratified subsonic and supersonic turbulence. At $r< \Rsonic$ where turbulence is supersonic the simulated VSF indices are $0.4-0.6$ consistent with Burgers. A steep index of $>0.6$ is apparent at $r>\Rsonic$.}
    \label{fig:slope_radial_profile} 
\end{figure*}

Fig.~\ref{fig:VSF_radii} further shows that the slopes of the VSFs below the break tend to be steeper at large radii than at small radii, in all three simulations. This trend is explored in Figure~\ref{fig:slope_radial_profile} which plots the slopes of the broken power-law fits versus radius, at the same snapshots and simulations used in Fig.~\ref{fig:VSF_radii} .
In all three simulations and radii the power law indices have values between $0$ and $1$, implying that the turbulent velocity increases with scale. At $r<R_{\rm sonic}$ where the turbulence is supersonic, the power law indices have a value of $0.4-0.6$, similar to the Burgers power law index of $0.5$ expected in supersonic turbulence. 
At $r>R_{\rm sonic}$ where the turbulence is subsonic, power law indices are higher than in the supersonic case, reaching $0.7-0.8$. These values are also higher than the Kolmogorov index of $1/3$ expected in homogeneous subsonic turbulence. 

\section{Discussion}\label{sec:discussion}

\subsection{Comparison to previous CGM inflow solutions}

The classic picture for 1D CGM inflows discussed in \cite{birnboim03} indicates a transition from free-falling `cold flows' at low halo masses, below a threshold in the range $10^{11}-10^{12}\msun$, to hot, cooling-limited inflows above the mass threshold. 
Our results, \changed{which account for the tendency of inflows to become turbulent \citep{Robertson12}, indicate that the low-mass regime is more accurately described as a turbulence dissipation-limited cold inflow rather than free-fall.} In this regime turbulence provides the bulk of support against gravity, turbulent velocities are comparable to the virial velocity, and turbulence Mach numbers substantially exceed unity. 
Furthermore, at intermediate halo masses a hybrid solution is expected, with a hot thermal pressure-dominated inflow in the outer CGM and a cool and turbulent inflow in the inner CGM, replacing the non-turbulent hybrid solution discussed in \cite{stern20}. 
The two parts of the hybrid solution are connected by a sonic radius which location generally decreases with \changed{increasing halo mass and with decreasing CGM mass relative to the halo baryon budget (see Figs.~\ref{fig:CGM_phases} -- \ref{fig:sim_img}). } 

Our assumption of quasi-spherical outer boundary conditions is justifiable for halo masses below the characteristic halo mass scale forming at a given epoch $M_{\rm grav}$, since such halos typically reside in filaments with size comparable to or larger than $\Rvir$, where  $M_{\rm grav}(z=2)\sim10^{10.5}\msun$, $M_{\rm grav}(z=1)\sim10^{12}\msun$, and $M_{\rm grav}(z=0)\sim10^{13}\msun$ \citep{Dekel06}. Above this mass scale halos are typically fed by streams that are narrow relative to the halo size, and thus the accretion geometry near $\Rvir$ significantly deviates from spherical \citep{Dekel06}. However, the extent to which filaments retain their identity within the halo rather than dissipate into the volume-filling CGM phase is an open question and an active field of study \citep[e.g.][]{Mandelker20}.  If filaments dissipate, the quasi-spherical accretion picture discussed in this work may also be applicable at halo masses above $M_{\rm grav}(z)$. 

We further note that studies of the interaction of filaments with their surroundings have assumed that the volume-filling phase is hot and quasi-static \citep[e.g.][]{Mandelker16,Yao25}. Our results suggest that it would be useful to examine also the interaction of cosmic filaments with a volume-filling medium that is cool and turbulent.

\subsection{Galaxy feedback and other unmodeled processes}\label{sec:feedback}

Previous studies of CGM turbulence have attributed turbulence-driving to the interaction of inflows with galaxy outflows and jets \citep{Fielding17,Buie18,Buie20,Buie22,Lochhaas20,Mohapatra22,Koplitz23,Pandya23} or to the hydrodynamical interaction of the cool and hot CGM phases \citep[e.g.,][]{Fielding20,Faerman25,Wibking25}. This common picture implies that turbulence properties, such as driving scale, turbulence velocity and Mach number, depend on the uncertain properties of galaxy feedback. The present work helps in disentangling the different turbulence-driving mechanisms by studying how turbulence is driven when only inflows are present. Our results thus provide a baseline for assessing the effect of feedback in simulations and observations.

The kinetic, thermal, and chemical effects of galaxy feedback can change CGM turbulence properties relative to our inflow-only solutions\footnote{Potential effects of cosmic ray feedback on the CGM are not discussed here. See, e.g., \cite{Ji20,Ji21,Butsky22,RodriguezMontero24,Hopkins25}.}. 
\changed{Feedback can change the cooling time of virial temperature gas $\tcools$ by altering the CGM density or metallicity.
These effects will change $\Rsonic$ and thus the threshold mass/radius for the transition between the cool and hot regimes. The magnitude of these effects can be estimated from eqn.~(\ref{eq:r_sonic}) and Fig.~\ref{fig:CGM_phases}, which demonstrate that $\Rsonic$ depends linearly on the CGM density and cooling rate. For example, in a halo with CGM mass fraction $\fCGM=0.2$  the threshold halo mass threshold where $\Rsonic$ exceeds $\Rcirc$ is $2.5\cdot10^{11}\msun$ at $z=0$ (see Fig.~\ref{fig:CGM_phases}), implying that turbulence-dominated CGM could be limited to dwarf-scale halos in the local Universe. This sensitivity of the mass threshold to CGM density and metallicity is most pronounced at $z\sim0$, due to the shape of the cooling function and the increase of $\Tvir$ with redshift at fixed halo mass (see fig.~9 in \citealt{stern20}).  }

\changed{Galaxy feedback can also inject energy into the CGM and offset cooling losses, effectively increasing the cooling time. One may thus presume that injecting heat could alter an inflow from the turbulence-dominated regime which requires rapid cooling to the thermal energy-dominated regime which requires cooling is relatively slow. We argue here that this is unlikely.} 
First, in many models galactic winds have velocities $\sim\vc$ \citep[e.g.,][]{Murray05,muratov15,Thompson16, anglesalcazar17b,Pandya21,Thompson24} which do not change the CGM specific energy from its no-feedback value of $\sim k\Tvir$. Such winds do not significantly alter the cooling times of CGM shocks, so a turbulence-dominated CGM where shocks are highly radiative will  remain so even when such winds are considered.  Alternatively, feedback in the form of outflows or jets could have a high specific energy $\gg k\Tvir$, causing the cooling time to significantly exceed the cooling time of virial temperature gas. In this case, however, the CGM would tend to become unbound, so the CGM would expand and adiabatically cool on a timescale $\lesssim\tff$, as can be seen for example in the idealized CGM simulations in \cite{Fielding17}. In between outflow episodes the CGM would cool down and again be turbulence dominated. 
Thus, only continuous injection of high specific energy outflows may be able to keep the CGM at a temperature $\gg\Tvir$, and thus produce a thermal energy-dominated CGM despite that the cooling time of $\sim\Tvir$ gas is smaller than $\tff$.  

\changed{Additional gravity-based turbulence stirring mechanisms, including stirring by dark matter subhalos, mergers, or a time-dependent gravitational potential, are unlikely to substantially increase the turbulent velocity above $\sigmat\approx\vc$. Thus, our inflow-only prediction of $\sigmat\approx\vc$ in halos with mass $\lesssim 10^{12}\msun$ is unlikely to change when these additional effects are taken into account. In a companion paper \cite{Kakoly25} we show that in the FIRE cosmological zoom simulations, where these effects are accounted for, we  indeed find $\sigmat\approx\vc$ in inner CGM of $\lesssim 10^{12}\msun$ halos as predicted by our inflow-only solutions. }

In halos with a mass above $10^{12}\msun$ where the inner CGM is dominated by thermal energy, the predicted turbulent velocity due to accretion is low relative to $\vc$ (Fig.~\ref{fig:sigma_t_M_vir}). 
Thus, it is plausible that other turbulence stirring mechanisms dominate over accretion in these halos, in contrast to lower mass halos where accretion alone already implies $\sigmat\approx\vc$. 

We note also that accounting for turbulence in the inflow will affect the properties of winds driven by feedback. The supersonic turbulence expected in low-mass halos  will allow galaxy outflows to reach larger radii than expected if the CGM is not turbulent, due to the formation of paths of least resistance via which the outflow can propagate, as discussed in the context of the ISM in \cite{Thompson16}.
This maximum radius of galaxy winds which fall back onto the galaxy has been called the `recycling' radius of the wind, and affects the mass and metal evolution of galaxies and their CGM \citep{Angles2017}. Turbulent accretion could also drive turbulence in the ISM, and thus could increase star formation burstiness and the strength of outflows (see discussions in \citealt{Fielding18} and \citealt{Stern21A}).

\subsection{Accretion rate and angular momentum coherence}\label{sec:accretion}
Accretion rates onto galaxies are an important quantity in galaxy evolution since they set star formation rates \citep[e.g.,][]{dekelnature09,FaucherGiguere11,Lilly13}.
Our inclusion of turbulence in CGM inflow solutions decreases the inflow speed from the free-fall velocity to the dissipation regulated velocity of $\uin=r/\tdiss$ (eqn.~\ref{eq:ur cool solution}), in $\lesssim10^{12}\msun$ halos where turbulence dominates (Fig.~\ref{fig:velocity_radial_profiles}).  This is a modest decrease of a factor of $\approx2$, and the reduction in the accretion rate may be even lower if the accretion rate is limited by the supply from the large-scale structure rather than by inflow speeds. A similar modest decrease by a factor of $\lesssim \sqrt{2}$ in accretion rates due to CGM turbulence was deduced by \cite{Pandya23} in their semi-analytic model. In more massive halos where the inflow is cooling-limited, our inclusion of turbulence has a negligible affect on inflow speeds. 

A more significant effect of adding turbulence to CGM inflow solutions is its implications for the angular momentum coherence of gas accreting onto the galaxy. \cite{Hafen22} argued that a narrow angular momentum distribution in accreting gas could be necessary for the formation of thin disk galaxies commonly seen in the low-redshift Universe \citep[e.g.,][]{kassin12,Tiley21}. They demonstrated that such a narrow angular momentum is indeed achieved in $\sim L^\star$ FIRE galaxies at $z\sim0$ which have $\Rsonic<\Rcirc$, where $\Rcirc$ is the circularization radius at which accretion onto the galaxy occurs (see Fig.~\ref{fig:CGM_phases}).
Our results provide physical intuition for the \citeauthor{Hafen22} result. The broadness of the angular momentum distribution in accreting gas can be quantified via 
\begin{equation}
\label{eq:ang mom coherence}
\frac{\sigma_j}{\langle j \rangle}\approx\sqrt{\frac{2}{3}}\frac{\sigmat}{\vc}~,
\end{equation}
where $\langle j \rangle$ and $\sigma_j$ are the mean and dispersion in the specific angular momentum of accreting gas. The approximation in eqn.~(\ref{eq:ang mom coherence}) is a result of \changed{$\langle j \rangle=\vc\Rcirc$ at $\Rcirc$, and of $\sigma_j^2\approx (2/3)\sigmat^2\Rcirc^2$} since tangential modes of turbulence create a dispersion in the angular momentum vector. Equation~(\ref{eq:ang mom coherence}) thus implies that in low-mass halos where accretion is in the `cold mode' and $\sigmat\approx\vc$, the angular momentum distribution is necessarily broad with $\sigma_j/\langle j \rangle\approx  0.8$. In contrast, in more massive halos where $\sigmat^2/\vc^2\ll1$ (Figs.~\ref{fig:sigma_t_M_vir} -- \ref{fig:velocities_cool2ff}) and the hot mode dominates, the angular momentum distribution is narrow with $\sigma_j/\langle j \rangle\ll1$. 
This distinction between accretion in the hot and cold modes explains the association of accretion mode and angular momentum coherence found by \cite{Hafen22} in the FIRE simulations.

\section{Summary}\label{sec:summary}

In this work we derive quasi-spherical solutions for turbulent CGM inflows, based on the framework for accretion-driven turbulence developed by \cite{Robertson12} and the cooling flow solutions discussed in \cite{stern19,stern20}. We focus on the physics of radiating CGM inflows in a constant, spherical dark matter potential, neglecting galaxy feedback and other complications such as satellites and mergers.

We show that turbulent CGM inflow solutions are mainly characterized by their sonic radius $\Rsonic$. Within $\Rsonic$ the inflow is predominantly cool with supersonic turbulence and inflow velocities, \changed{and the inflow is `dissipation-limited' --- regulated by the rate at which turbulence dissipates. Outside $\Rsonic$ the inflow is predominantly hot with subsonic turbulence and inflow velocities, and the flow is `cooling-limited' --- regulated by the radiative cooling rate as in a classic ICM cooling flow. The location of $\Rsonic$ relative to $\Rvir$ generally decreases with increasing mass, such that in $\gtrsim10^{12}\msun$ halos CGM inflows are entirely subsonic, while at lower masses the CGM is supersonic at least at inner halo radii (Figs.~\ref{fig:CGM_phases} -- \ref{fig:sim_img}). This threshold halo mass decreases if the CGM mass has been strongly depleted by feedback (see Fig.~\ref{fig:CGM_phases}).}  

Using idealized 3D hydrodynamic simulations, we demonstrate that average CGM properties indeed converge to our analytic solutions. 
We use the simulations also to study gas density distributions, velocity structure functions (VSFs), and turbulence anisotropy.   
Our main conclusions can be summarized as follows:

\begin{enumerate}
  \item `Adiabatic heating' of turbulence in CGM inflows causes the turbulent velocity to increase inwards, until saturating at $\sigmat=A\uin$ where $\uin$ is the radial inflow velocity and $A\approx1.5-3$ (Fig.~\ref{fig:sigma_t_over_u_r}). Saturation occurs when the inflow time equals the turbulence dissipation time.
  \item The mean sound speed $\langle c_{\rm s}\rangle$ and turbulence Mach number at a given radius are determined by whether the cooling time of virial temperature gas $\tcools$ (eqn.~\ref{eq:t_cool}) is larger or smaller than the free-fall time $\tff$. When $\tcools>\tff$ (inner CGM at $\Mhalo\gtrsim10^{12}\msun$ and outer CGM at $\gtrsim10^{11}\msun$) the CGM is predominantly hot and turbulence is subsonic, similar to the ICM. In contrast when $\tcools<\tff$ the CGM is predominantly cool and turbulence is supersonic (Fig.~\ref{fig:velocities_cool2ff}). 
  \item At halo masses and radii for which turbulence is supersonic, turbulence provides the dominant mode of support against gravity, with $\sigmat\approx\vc\gg\langle c_{\rm s}\rangle$. Inflow velocities in this regime are somewhat lower than free-fall, with $\uin\approx\vc/2$ (Fig.~\ref{fig:velocity_radial_profiles}). 
   \item Density distributions are narrow in the subsonic regime and wide in the supersonic regime, as seen in simulations of isothermal turbulent boxes. A bi-modal distribution associated with the two thermal phases is evident only in intermediate mass halos where the turbulent Mach number is of order unity (Fig.~\ref{fig:density_PDF}). 
  \item By measuring VSFs in radial shells we find the expected `Burgers' slope of $0.5$ in the supersonic turbulence regime. In the subsonic regime we find a high VSF slope of $0.7$ (Fig.~\ref{fig:slope_radial_profile}). 

  \item \changed{We argue that our $\sigmat\approx\vc$ prediction for rapidly-cooling CGM likely holds also when accounting for additional physical processes, such as ongoing feedback, mergers, and turbulence stirring by subhalos (section~\ref{sec:feedback}). Indeed, FIRE cosmological zoom simulations which account for stellar feedback and dark matter halo evolution also exhibit $\sigmat\approx\vc$, as shown in other papers by our group \citep{Gurvich23,Kakoly25,Sun25}.}
\end{enumerate}

Our result that turbulence energy in the CGM can dominate over thermal energy implies that the  interaction between gas elements is dominated by ram pressure rather than thermal pressure, yielding Burgers-like velocity structure functions (Fig.~\ref{fig:slope_radial_profile}) and a wide lognormal density distribution  (Fig.~\ref{fig:density_PDF}).
Such a turbulence-dominated picture for the inner CGM of $\lesssim L^\star$ galaxies qualitatively differs from the common `multi-phase' picture where cool localized `clouds'  are embedded in a hot and quasi-static volume-filling medium. Discerning the properties of such turbulence-dominated CGM is thus imperative for interpreting observations from this regime and for understanding its implications for galaxy evolution. 
In other papers by our group we discuss predictions of turbulence-dominated CGM for UV absorber equivalent widths \citep{Kakoly25}, for high redshift star formation \citep{Sun25}, for the coherence of accreting gas \citep[see also section~\ref{sec:accretion}]{Hafen22}, and for high redshift DLA observations \citep{Stern21B}. 
Additional implications of turbulence-dominated CGM worth exploring in future work include  absorber coherence scales \citep[e.g.][]{Afruni23}, \changed{comparison of predicted velocity structure functions with observations \citep{Rauch01, Chen23}}, and the overlap in velocity space of ions with widely different ionization levels \citep{Werk16}.

\section*{Acknowledgements}
The Authors thank Yuval Birnboim for insightful discussions related to this paper. 
RG and JS were supported by the Israel Science Foundation (grant No.\ 2584/21) and by a grant from the United States-Israel Binational Science Foundation (BSF), Jerusalem, Israel.
CAFG was supported by NSF through grants AST-2108230 and AST-2307327; by NASA through grants 80NSSC22k0809, 80NSSC22K1124 and 80NSSC24K1224; by STScI through grant JWST-AR-03252.001-A; and by BSF through grant \#2024262. 

\section*{Data Availability}
The simulation data underlying this article will be shared on reasonable request to the corresponding authors. A public version of the GIZMO simulation code is available at \hyperlink{https://github.com/pfhopkins/gizmo-public}{https://github.com/pfhopkins/gizmo-public}.



\bibliographystyle{mnras}
\bibliography{References} 




\titleformat{\section}
  {\normalfont\bfseries}
  {Appendix \thesection:}{1em}{}

\appendix

\section{Adiabatic heating}\label{app:adiabatic heating}

The adiabatic heating relation (eqn.~\ref{eq:heating}) from \cite{Robertson12}  can be understood as follows. 
Assume a gas fluid element in a homologously contracting medium, traveling from point $1$ with velocity $\vec{u}_\mathrm{1}$ towards point $2$ at a distance $\vec{\dr}$, where $1$ and $2$ are comoving with the flow. The velocity between the points due to the contraction is proportional to the contraction rate and the distance between the points:
\begin{equation}
    \vec{u}_{12}=\frac{\dot{r}}{r}\vec{\dr}=\frac{\ur}{r}\vec{\dr}.
	\label{eq:expansion_vel}
\end{equation}
When the fluid element reaches point $2$ it has not been accelerated, so the velocity remains $\vec{u}_1=\vec{u}_2+\vec{u}_{12}$ where $\vec{u}_2$ is the velocity of the fluid element relative to point $2$. The change of relative velocity during a time $\delta t$ is then:
\begin{equation}
    \frac{\delta \vec{u}}{\delta t}=\frac{\vec{u}_\mathrm{2}-\vec{u}_\mathrm{1}}{\delta t}=-\frac{\vec{u}_\mathrm{21}}{\delta t}=-\frac{\ur}{r}\frac{\vec{\dr}}{\delta t}.
	\label{eq:adiabatic_derive}
\end{equation}
On small $\delta t$ one can approximate $\vec{\dr}=\vec{u}_{1}\,\delta t$. Thus:
\begin{equation}
    \frac{\delta \vec{u}}{\delta t}=-\vec{u}_\mathrm{1}\frac{\ur}{r}.
	\label{eq:adiabatic_term}
\end{equation}
This relation is true for each fluid element separately, so the turbulent velocity $\sigmat$ which is the average of velocity of random motions will also obey the same equation, which explains eqn.~(\ref{eq:heating}). 

\section{VSF of cool, supersonic inflows}\label{app:VSF}

In this appendix we derive eqn.~(\ref{eq: VSF with heating}), an estimate of the first-order VSF in a supersonic, turbulence-dominated inflow. 

For scales $\dr$ in the inertial range, the VSF is expected to follow a Burgers spectrum ($\du\propto \dr^{1/2}$) in isotropic and homogeneous supersonic turbulence.  
\cite{Matzner07} showed that this spectrum 
can be derived by assuming a downward cascade of linear momentum between scales $\dr$, akin to the cascade of energy used to derive the Kolmogrov spectrum. This cascade can be derived from the equation 
\begin{equation}\label{eq:momentum l}
     -\frac{d}{d\ln\dr}\left(\eta\frac{(\du)^2}{\dr}\right) = 0 ~,
\end{equation}
 where the term in parenthesis is the momentum transfer rate per unit mass for an eddy turnover time of $\eta\du/\dr$, and $\eta$ is the dissipation constant (eqn.~\ref{eq:dissipation}). The negative sign denotes explicitly that the cascade is downward in $\dr$. 
 
Adiabatic heating is scale-independent (see appendix~\ref{app:adiabatic heating}), so will modify eqn.~(\ref{eq:momentum l}) by adding a source term on all scales akin to eqn.~(\ref{eq:heating}): 
\begin{equation}\label{eq:momentum l modified}
    -\frac{d}{d\ln\dr}\left(\eta\frac{(\du)^2}{\dr}\right) = \frac{\du}{r/\uin}  ~,
\end{equation}
where we now assume that $\du$ is  a function of both $\dr$ and $r$, representing the VSF of a shell of accreting gas measured at radius $r$. 
Eqn.~(\ref{eq:momentum l modified}) can be solved using $\uin=\eta\sigmat$ (eqn.~\ref{eq:cool_solution}), which yields:
 \begin{equation}\label{eq: VSF with heating app}
\du(\dr,r) = \sigmat\left[C\left(\frac{\ell}{r}\right)^{1/2} - \frac{\ell}{r}\right] ~,
\end{equation}
where $C$ is an integration constant which satisfies $\du(r,r) = (C-1)\sigmat(r)$, so we expect $C\approx2$.  
Note that our use of the cascade term in eqn.~(\ref{eq:momentum l modified}) is inaccurate at scales $\dr\sim r$ since there are no larger scales to cascade from. This inaccuracy makes only a small difference since the implied cascade from scales $\dr>r$ is small. 

For eqn.~(\ref{eq: VSF with heating}) in section~\ref{sec:VSF analytic} we used eqn.~(\ref{eq: VSF with heating app}) with $C=5/3$, which provides a reasonable fit to the simulated VSF at small $\ell$ (see Fig.~\ref{fig:VSF_example}).

\section{Resolution study}\label{app:resolution}

\begin{figure}
	
 \centering
	\includegraphics[width=\columnwidth]{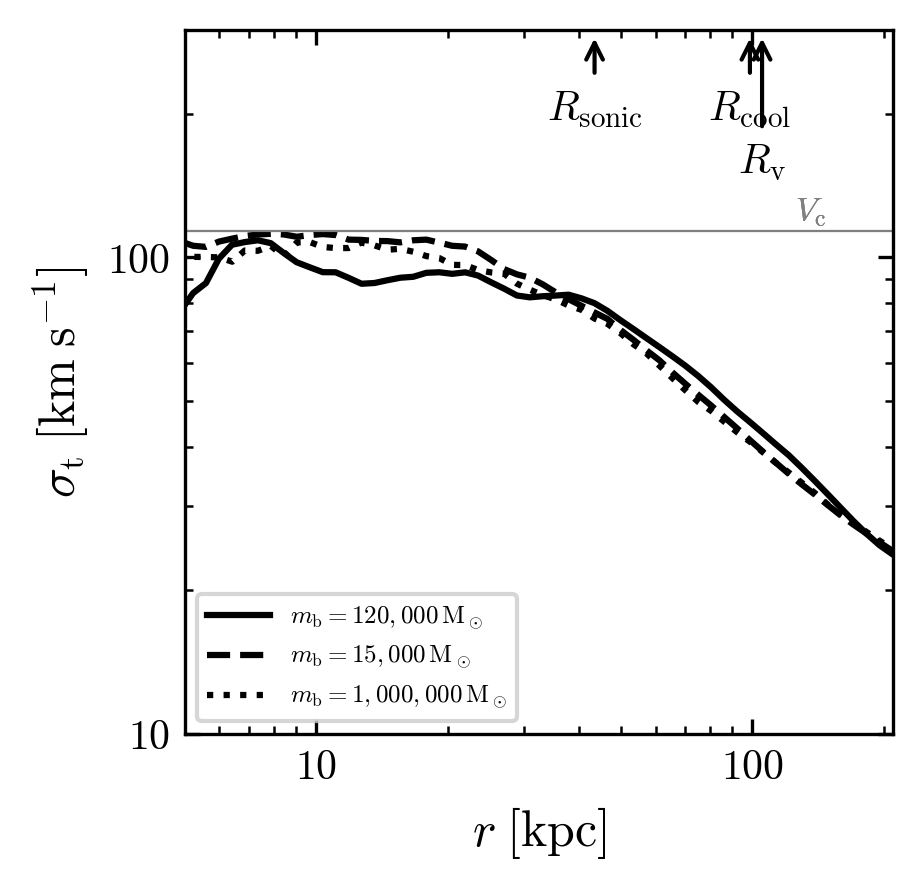}

    \vspace{-0.35cm}
    \caption{\changed{Turbulent velocity profiles of CGM accretion flows in a $10^{11.5}\msun$ halo at $z=1$, in simulations with different gas mass resolutions.  
    Profiles are averaged over snapshots after a quasi-steady flow has formed, as in Figs.~\ref{fig:velocity_radial_profiles} -- \ref{fig:different_ics_vel}. The similarity of the  $\sigmat$ profiles among the three simulations suggests they are independent of resolution at the range probed. }}
    \label{fig:different_res_vel}
\end{figure}

\changed{
Figure~\ref{fig:different_res_vel} explores how our conclusion that $\sigmat\approx\vc$ at inner CGM radii depends on simulation resolution. 
The figure compares the $\sigmat$ profile in the fiducial simulation (`m115z1', solid line), with a simulation with $\times8$ higher gas mass resolution (`m115z1\_d', dashed line) and a simulation with $\times8$ lower gas mass resolution (`m115z1\_e', dotted line). We plot the mean profiles when the flow is quasi-steady, at the same simulation times as shown in Fig.~\ref{fig:velocity_radial_profiles} -- \ref{fig:different_ics_vel}. 
In all three simulations the turbulent velocity profiles increase inward and reach $\sigmat\approx\vc$ at small radii, suggesting that this result is independent of resolution at the range probed.
}

\label{lastpage}
\end{document}